\documentclass[journal,10pt]{IEEEtran}

\usepackage{algorithm} 
\usepackage{algorithmic} 
\usepackage{multirow} 
\usepackage{amsmath}
\usepackage{xcolor}
\usepackage{amsmath}
\usepackage{stfloats} 
\usepackage{amssymb}
\usepackage{amsmath}
\usepackage{cite}
\usepackage{url}
\usepackage{xcolor}
\usepackage{cite,graphicx,amsmath,amssymb}
\usepackage{subfigure}
\usepackage{citesort}
\usepackage{fancyhdr}
\usepackage{mdwmath}
\usepackage{mdwtab}
\usepackage{caption}
\usepackage{amsthm}

\usepackage{subfigure}

\usepackage{setspace}

\newtheorem{remark}{Remark}
\newtheorem{theorem}{Theorem}

\newtheorem{lemma}{Lemma}

\newtheorem{corollary}{Corollary}

\makeatletter
\def\ScaleIfNeeded{%
\ifdim\Gin@nat@width>\linewidth \linewidth \else \Gin@nat@width
\fi } \makeatother

\begin{document}

\title{Caching Placement and Resource Allocation for Cache-Enabling UAV NOMA Networks}

\author{

Tiankui~Zhang,~\IEEEmembership{Senior Member,~IEEE,}
        Ziduan~Wang,
        Yuanwei~Liu,~\IEEEmembership{Senior Member,~IEEE,}
        Wenjun~Xu,~\IEEEmembership{Senior Member,~IEEE}
        and Arumugam Nallanathan,~\IEEEmembership{Fellow,~IEEE}

\thanks{
This work was supported by National Natural Science Foundation of China under Grants 61971060 and 61502046.
}
\thanks{Tiankui Zhang, Ziduan~Wang and Wenjun~Xu are with the School of Information and Communication Engineering, Beijing University of Posts and Telecommunications, Beijing 100876, China (e-mail: \{zhangtiankui, wangziduan, wjxu\}@bupt.edu.cn). }
\thanks{Yuanwei Liu and Arumugam Nallanathan are with the School of Electronic Engineering and Computer Science, Queen Mary University of London, London E1 4NS, U.K. (e-mail:  \{yuanwei.liu, a.nallanathan\}@qmul.ac.uk).}
}

\maketitle
\begin{abstract}
This article investigates the cache-enabling unmanned aerial vehicle (UAV) cellular networks with massive access capability supported by non-orthogonal multiple access (NOMA). The delivery of a large volume of multimedia contents for ground users is assisted by a mobile UAV base station, which caches some popular contents for wireless backhaul link traffic offloading.
In cache-enabling UAV NOMA networks, the caching placement of content caching phase and radio resource allocation of content delivery phase are crucial for network performance.
To cope with the dynamic UAV locations and content requests in practical scenarios, we formulate the long-term caching placement and resource allocation optimization problem for content delivery delay minimization as a Markov decision process (MDP). The UAV acts as an agent to take actions for caching placement and resource allocation, which includes the user scheduling of content requests and the power allocation of NOMA users. In order to tackle the MDP, we propose a Q-learning based caching placement and resource allocation algorithm, where the UAV learns and selects action with \emph{soft ${\varepsilon}$-greedy} strategy to search for the optimal match between actions and states. Since the action-state table size of Q-learning grows with the number of states in the dynamic networks, we propose a function approximation based algorithm with combination of stochastic gradient descent and deep neural networks, which is suitable for large-scale networks. Finally, the numerical results show that the proposed algorithms provide considerable performance compared to benchmark algorithms, and obtain a trade-off between network performance and calculation complexity.
\end{abstract}

\providecommand{\keywords}[1]{\textbf{\textit{Index terms---}} #1}


\section{Introduction}
With the explosion of massive multimedia applications and the continuous growth of mobile data traffic, wireless communication faces the problem of limited resources.
In order to effectively meet the increasing user demand for high data rate and low access delay, many works~\cite{Cui2018Multi,Liu2018UAV,7572068,ZhangTk20TII,R18907440} have paid attention to  wireless connectivity from the sky with unmanned aerial vehicles (UAVs). UAVs, also known as remotely piloted aircraft systems (RPAS) or drones, are small pilotless aircrafts that are rapidly deployable for complementing terrestrial communications~\cite{Cui2018Multi}.
Promising scenarios for UAV communications can be as follows: establishing temporal communication infrastructure during natural disasters, offloading traffic for dense cellular networks, data collection for supporting Internet of Things~(IoT)~\cite{Liu2018UAV}, and mobile edge computing server for supporting IoT~\cite{ZhangTk20TII}.

With the rapid growth of UAV-assisted cellular networks, UAVs perform diverse roles, including radio access nodes, base stations (BSs)~\cite{R18907440,Hou2018Multiple,8629316,R18677286,TWC8685130,access8809879} and relays~\cite{TWC8424236}. 
In UAV-assisted cellular networks, the data rate is limited by both the radio access links and the wireless backhaul links.
For radio access links, multi-users can be served with the same time region and frequency band based on non-orthogonal multiple access (NOMA), which has received remarkable attention~\cite{7812773,7982794,R18936405}.
In~\cite{7812773}, the performance of NOMA in large-scale networks has been investigated with stochastic geometry theory. The resource allocation of NOMA heterogeneous networks (HetNets) has been studied in~\cite{7982794}.
In~\cite{R18936405}, the pair-wise error probability (PEP) performance of different detectors in multiple-input-multiple-output (MIMO) NOMA system has been analyzed, which has also been minimized by the proposed two kinds of user selection methods.
Recently, NOMA has been exploited as an effective method to enhance the access capability of UAV-assisted cellular networks~\cite{Hou2018Multiple,8629316,R18677286,TWC8685130,access8809879}.
To mitigate the traffic load of the backhaul links, edge caching has been studied~\cite{8382257,8002665}.
In~\cite{8382257}, the bandwidth allocation and caching placement have been jointly optimized in HetNets.
In~\cite{8002665}, multi-tier collaborative caching framework in HetNets has been studied to maximize the network capacity.
Caching popular contents at UAVs has been regarded as an effective emerging method to alleviate the backhaul congestion and reduce latency in UAV-assisted cellular networks~\cite{ZhangTK20TWC,8614433,7875131,8254370,8626132}.
The cache-enabling UAV NOMA network is a promising framework for high data rate and low transmission latency in numerous multimedia contents distribution scenario.
\subsection{Related Works}
Recently, there has been some researches on UAV NOMA cellular networks~\cite{Hou2018Multiple,8629316,R18677286,TWC8685130,access8809879}. Based on stochastic geometry theory, the 3-Dimension UAV framework for providing wireless service to randomly roaming NOMA users has been studies in~\cite{Hou2018Multiple}. In the UAV NOMA cellular networks, a cooperation mechanism between a UAV and a macro base station has been proposed in~\cite{8629316}, which maximizes the sum rate of users by the UAV trajectory and NOMA precoding joint optimization.
In~\cite{R18677286}, a hybrid transmission strategy based on VP and NOMA (VP-NOMA) has been proposed, which minimizes total transmit power for certain quality of service (QoS) requirements by designing a beamforming matrix with the power allocation strategy.
The UAV trajectory and resource allocation have been jointly optimized for maximizing the minimum data rate of ground users in
~\cite{TWC8685130}. In~\cite{access8809879}, the trajectory, task data, and computing resource allocation have been joinly optimized to minimize the largest energy consumption among UAVs.

Moreover, caching at UAV~\cite{ZhangTK20TWC,8614433,7875131,8254370,8626132} has attracted increasing attention recently to relieve the pressure on the wireless backhaul links.
The resource allocation in cache-enabling UAV-assisted cellular networks has been considered in~\cite{ZhangTK20TWC}, where a joint optimization problem of UAV deployment, caching placement and user association has been solved to maximize the quality of experience (QoE) of users.
In~\cite{8614433}, user association, spectrum allocation, and content caching have been jointly optimized by a liquid state machine (LSM) based algorithm.
In~\cite{7875131}, user association, UAV location, and caching placement have been jointly optimized
to maximize the users' QoE while minimizing the transmit power used by the UAVs.
The cache-enabling UAV assisted secure transmission for scalable videos in hyper-dense networks has been studied in~\cite{8254370}, and a distributed algorithm has been proposed to manage the interference by cooperatively generating the precoding matrices of cache-enabling UAVs. In~\cite{8626132}, an optimization problem has been formulated to maximize the security of UAV-relayed wireless networks with caching by jointly adjusting the UAV trajectory and time scheduling.

The problem of resource allocation with dynamic networks has been studied in~\cite{7006718,7120184}. In~\cite{7006718}, a dynamic programming approach for heterogeneous networks (HetNets) has been designed, where communicating nodes have been efficiently matched and radio resources have been assigned in an interference-aware manner.
The energy harvesting downlink MIMO systems have been studied in~\cite{7120184}, where an online resource allocation algorithm has been proposed to maximized the sum rate.
Some works on the resource allocation of UAV-assisted cellular networks have been studied~\cite{R48039268,s28438896}.
However, only a few existing works have concentrated on the dynamic resource allocation of UAV-assisted cellular networks~\cite{R48254367,R49055054}.
A real-time access points provision algorithm has been developed in~\cite{R48254367}, where UAV-mounted cloudlets are assumed to carry out adaptive recommendation in a distributed manner so as to reduce computing and traffic load.
In~\cite{R49055054}, sensing and transmission protocol, UAV trajectory design,
and radio resource management in U2X communication have been jointly optimized to maximize the average number of valid data transmissions.
\subsection{Motivation and Contribution}
As mentioned above, the caching placement and resource allocation optimization problems have been considered in cache-enabling UAV NOMA networks. However, the optimization scenarios of the current studies are most likely to be static, and rarely consider the dynamic environment including the UAV movement and content request varying. Due to the moving characteristic of UAVs~\cite{6262501}, the efficiency of caching placement and resource allocation may be improved by considering a long-term optimization problem.
To fulfill this gap, this article studies the caching placement and resource allocation in cache-enabling UAV NOMA networks with dynamic UAV locations and content requests. The dynamic optimization problem for caching placement of a UAV, user scheduling of content requests, as well as power allocation of NOMA users is formulated. The caching placement and resource allocation process is modeled as a Markov decision process (MDP), which is solved by reinforcement learning. Moreover, a function approximation based algorithm is proposed to deal with the dynamic optimization problem in large-scale networks. Our contributions are summarized as follows:
\begin{itemize}
	\item We propose a framework of cache-enabling UAV NOMA cellular networks for content delivery of ground users in a hotspot area. We define the long-term sum delay of users as the content delivery cost of downlink UAV NOMA cellular networks. We formulate an optimization problem to minimize content delivery delay by jointly optimizing the caching placement of a UAV, user scheduling of content requests, and the power allocation of NOMA users.

	\item We transform the original proposed optimization problem to a MDP based problem and define the dynamic states of UAV movement and content request varying, in which the UAV performs as the agent. Since the instantaneous content delivery delay depends on the current state and action according to the property of Markov chain, we develop a Q-learning based content placement and resource allocation algorithm for solving the MDP based problem. Furthermore, to deal with the high complexity of an action-state table (Q-table) in large-scale networks, we propose a function approximation based caching placement and resource allocation algorithm, which can obtain the near-optimal solution according to the output of the function rather than searching a huge action space.

	\item We provide simulation results to validate the performance of the proposed caching placement and resource allocation algorithms compared with the benchmark algorithms. The simulation results demonstrate that the proposed Q-learning based algorithm gets a trade-off between network performance and computation complexity. Meanwhile, the proposed function approximation based algorithm obtains a considerable network performance without the complexity ergodic search in the action space.
\end{itemize}

\subsection{Organization}
The rest of this article is organized as follows. In Section \uppercase\expandafter{\romannumeral2}, we present the system model and formulate the optimization problem for long-term content delivery delay minimization. In Section \uppercase\expandafter{\romannumeral3}, we propose reinforcement learning based algorithms for caching placement and resource allocation.
Simulation results are presented in Section \uppercase\expandafter{\romannumeral4},
which is followed by conclusion in Section \uppercase\expandafter{\romannumeral5}.

\begin{figure} [t!]
	\centering
	\includegraphics[width= 3.5in]{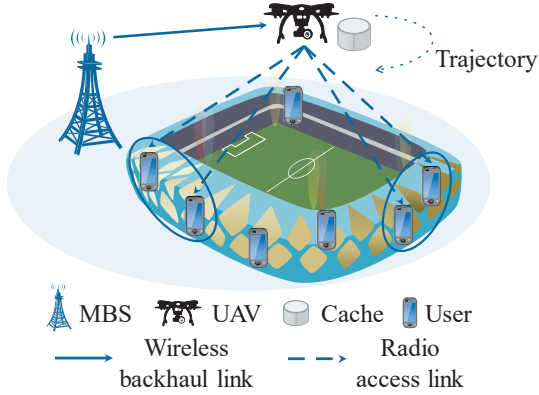}
	\caption{Cache-enabling UAV NOMA networks.}
		\label{fig1systemmodel}
\end{figure}
\section{System Model and Problem Formulation}
We consider the downlink transmission for $N$ users in a hotspot area covered by a ground macro base station (MBS) assisting by a mobile UAV base station. The UAV is connected to the MBS via wireless backhaul links, as shown in Fig.~\ref{fig1systemmodel}. We consider a dynamic scenario, which consists of a periodically-moving UAV and time-varying request for contents. We assume that the flight trajectory of the UAV is predetermined as in~\cite{s18624565} and the flight time is ${{T_v}}$.

We use the NOMA technique, including superposition coding (SC) technique at the UAV and serial interference cancel (SIC) technique at users. Two users form a group and we denote group set as ${{\cal G} = \{ 1,2, \cdot  \cdot  \cdot ,G\}}$. The UAV could serve more than one group at the same time slot. In each group, the user closer to the UAV is called as the near user (NU), and the other user is called as the far user (FU). Following the nearest near user and nearest far user (NNNF) NOMA users selection scheme proposed in~\cite{7445146}, we assume that the  nearest NU and nearest FU are assigned to a group. We define the mapping from users to group $g$ at time slot ${t}$ as ${\left( {{\varphi _{{\rm NU}g}}\left( t \right),{\varphi _{{\rm FU}g}}\left( t \right)} \right)}$. Table \ref{table1} provides a summary of the notations used hereinafter.

\begin{table}[t!]
\small
\caption{Notation}
\label{table1}

\begin{tabular}{|l|l|}
\hline
Notation                                                                                                                                       & Description                                                                                                              \\ \hline
${{\cal N}}$                                                                                 & Set of users                                                                                                             \\ \hline
${{T_v} }$                                                                                                                                     & Fly period of UAV                                                                                                        \\ \hline
${{\cal M}}$                                                                                                                                   & Set of contents                                                                                                          \\ \hline
${C_1}$                                                                                                                                        & Size of content                                                                                                          \\ \hline
${Z}$                                                                                                                                          & Cache capacity of UAV                                                                                                    \\ \hline
${{L_{n}}}$, ${{L_{\rm ma}}}$                                                                                                                  & Location of user $n$ and MBS                                                                                             \\ \hline
${{D_{{\rm B}m}}\left( t \right)}$                                                                                                               & Backhaul link delay of content $m$                                                                                       \\ \hline
${{q_m}}$                                                                                                                                      & Probability of users request for content $m$                                                                             \\ \hline
${{i_m}\left( t \right)}$                                                                                                                      & Proactive cache policy of content $m$                                                                                    \\ \hline
${{d_{\rm ma}}\left( t \right)}$                                                                                                               & Distance between MBS and UAV                                                                                             \\ \hline
${\Pr \left( {{g_{\rm maLoS}}\left( t \right)} \right)}$                                                                                       & LoS probability of backhaul link                                                                                         \\ \hline
${{h_g}\left( t \right)}$                                                                                                                      & Coefficient of power allocation of group $g$                                                                             \\ \hline
${{\cal G} }$                                                                                               & Set of user groups                                                                                                       \\ \hline
${\delta }$                                                                                                                                    & Length of time slot                                                                                                      \\ \hline
${{p_{\rm ma}}}$, ${{p_{{\rm uav}}}}$                                                                                                          & Power of MBS and UAV                                                                                                     \\ \hline
${\eta }$                                                                                                                                      & Zipf distribution Parameter                                                                                              \\ \hline
${{b_n}\left( t \right)}$                                                                                                                      & Response to user $n$ at time slot $t$                                                                                    \\ \hline
${{L_{\rm uav}}\left( t \right)}$                                                                                                              & Location of UAV at time slot $t$                                                                                         \\ \hline
${{D_{{\rm A}n}}(t)}$                                                                                                                            & Radio access link delay of user $n$                                                                                      \\ \hline
${{c_m}\left( t \right)}$                                                                                                                      & \begin{tabular}[c]{@{}l@{}}Cache situation of content $m$ in the\\  beginning of time slot $t$\end{tabular}              \\ \hline
\begin{tabular}[c]{@{}l@{}}${{d_{{\rm NU}g}}(t)}$, \\ ${{d_{{\rm FU}g}}(t)}$\end{tabular}                                                      & \begin{tabular}[c]{@{}l@{}}Distance between UAV and users of group \\ $g$ at time slot $t$\end{tabular}                  \\ \hline
${{\bar g_{\rm ma}}\left( t \right)}$                                                                                                          & Average path loss of backhaul link                                                                                       \\ \hline
${{r_{nm}}}$                                                                                                                                   & Request of user $n$ to content $m$                                                                                       \\ \hline
${{\mu _m}\left( t \right)}$                                                                                                                   & Content virtual queues in time slot $t$                                                                                  \\ \hline
\begin{tabular}[c]{@{}l@{}}${{R_{\rm B}}\left( t \right)}$,${{R_{{\rm NU}g}}\left( t \right)}$,\\ ${{R_{{\rm FU}g}}\left( t \right)}$\end{tabular} & \begin{tabular}[c]{@{}l@{}}Data rate of backhaul link, NU's and FU's\\  radio access links at time slot $t$\end{tabular} \\ \hline
\begin{tabular}[c]{@{}l@{}}${{\Gamma _{{\rm NU}g}}\left( t \right)}$, \\ ${{\Gamma _{{\rm FU}g}}\left( t \right)}$\end{tabular}                & \begin{tabular}[c]{@{}l@{}}SINR of radio access link of NU and \\ FU at time slot $t$\end{tabular}                       \\ \hline
\begin{tabular}[c]{@{}l@{}}${{\bar g_{{\rm NU}g}}\left( t \right)}$,\\ ${{\bar g_{{\rm FU}g}}\left( t \right)}$\end{tabular}                   & \begin{tabular}[c]{@{}l@{}}Average path loss of radio access link of \\ users in group $g$ at time slot $t$\end{tabular} \\ \hline
\begin{tabular}[c]{@{}l@{}}${{g_{\rm maLoS}}\left( t \right)}$, \\ ${{g_{\rm maNLoS}}\left( t \right)}$\end{tabular}                           & \begin{tabular}[c]{@{}l@{}}LoS and NLoS path loss of backhaul link\\  at time slot $t$\end{tabular}                      \\ \hline
\end{tabular}
\end{table}

\subsection{UAV Mobility Model}
As indicated above, the UAV flies on the trajectory with finite time period ${{T_v}}$. For the convenience of description, the flying duration ${{T_v}}$ is discretized into ${{T}}$ equal time slots, i.e., ${{T_v} = T\delta }$, where ${\delta}$ is the length of each time slot. We assume moving speed of the UAV is ${v}$, and the maximum moving distance of the UAV in each time slot is ${\delta v}$. Therefore, we can assume that the distance between the MBS and the UAV does not change during each time slot, so is the distance between the UAV and users. To simplify the problem, we focus on the stable flight process of the UAV at height $h$, ignoring the UAV's take-off and landing phases.
We use ${{L_{{\rm uav}}}\left( t \right):\left( {{x_{\rm uav}}\left( t \right),{y_{\rm uav}}\left( t \right),h} \right)}$ to express the observed location of the UAV at time slot $t$, and the location of user $n$ is defined as ${{L_{\rm n}}: ({x_{\rm n}},{y_{\rm n}})}$, which is subject to random distribution. We use ${{L_{\rm ma}}: ({x_{\rm ma}},{y_{\rm ma}})}$ to represent the location of the MBS.
We define the distance between the MBS and the UAV at time slot $t$ as ${{d_{\rm ma}}\left( t \right)}$, which is calculated as
\begin{equation}
{\begin{array}{l}
	{d_{\rm ma}}\left( t \right) = \sqrt {{{\left( {{x_{\rm ma}} - {x_{\rm uav}}\left( t \right)} \right)}^2} + {{\left( {{y_{\rm ma}} - {y_{\rm uav}}\left( t \right)} \right)}^2} + {h^2}}
	\end{array}}.
\end{equation}
Similarly, we obtain the distance from the NU and the FU of group $g$ to the UAV at time slot $t$, ${{d_{{\rm NU}g}}(t)}$ and ${{d_{{\rm FU}g}}(t)}$.
Since the location of the UAV is dynamic in the flying duration, the distances between the UAV and the MBS/users are time varying with time slot $t$.

\newcounter{TempqCnt}
\setcounter{TempqCnt}{\value{equation}}
\setcounter{equation}{4}
\begin{figure*}[hb]
	\hrulefill
	\begin{align}
	\label{7}
	{\Pr \left( {{g_{\rm maLoS}}\left( t \right)} \right) = \left\{ {\begin{array}{*{20}{c}}
			{1,{\rm{ }}if{\rm{ }}\sqrt {{d_{\rm ma}}{{\left( t \right)}^2} - {h^2}}  \le {d_o},}\\
			{\frac{{{d_o}}}{{\sqrt {{d_{\rm ma}}{{\left( t \right)}^2} - {h^2}} }} + \exp \left\{ {\left( {\frac{{ - \sqrt {{d_{\rm ma}}{{\left( t \right)}^2} - {h^2}} }}{{p_1}}} \right)\left( {1 - \frac{{{d_o}}}{{\sqrt {{d_{\rm ma}}{{\left( t \right)}^2} - {h^2}} }}} \right)} \right\},{\rm{ }}if{\rm{ }}\sqrt {{d_{\rm ma}}{{\left( t \right)}^2} - {h^2}}  > {d_o}.}
			\end{array}} \right.}
	\end{align}
\end{figure*}
\setcounter{equation}{\value{TempqCnt}}
\subsection{Content Request and Cache Model}
We denote the multimedia contents as a set ${{\cal M} = \{ 1, \cdot  \cdot  \cdot ,M\} }$.
At time slot $t$, we assume the probability that user $n$ requests content $m$ follows Zipf distribution, which is a conditional probability with user $n$ generating request at time slot $t$ as the condition, i.e. ${{{r_n}\left( t \right) = 1}}$, and can be calculated as
\begin{equation}
\label{zipf2}
{P\left( {{r_{nm}}\left( t \right) = 1|{r_n}\left( t \right) = 1} \right) = \frac{{\frac{1}{{{m^\eta }}}}}{{\sum\nolimits_{{j_1} = 1}^M {\frac{1}{{{j_1}^\eta }}} }}},
\end{equation}
where ${\eta }$ is the exponent of the Zipf distribution. Let ${{r_{nm}} \left( t \right) = 1}$ if user $n$ requests content $m$ at time slot $t$, otherwise ${{r_{nm}}\left( t \right) = 0}$. The user $n$ requests for at most one content at time slot $t$.
We assume that the preference of user to content is constant
over time, which can be extended to time varying user preference occasion. 	
In this model, the probability of user $n$'s request for content $m$ follows random average distribution over the time slots. The user $n$ requests for contents at time slot $t$ with the probability $P\left( {{r_n}\left( t \right) = 1} \right) = \frac{{{R_g}}}{N}$, where the request generating coefficient ${{R_g}}$ is constant over time. Then the probability that user $n$ requests content $m$ at time slot $t$ is given by ${P\left( {{r_{nm}}\left( t \right) = 1} \right) = P\left( {{r_{nm}}\left( t \right) = 1|{r_n}\left( t \right) = 1} \right)P\left( {{r_n}\left( t \right) = 1} \right)}$.
%
\begin{remark}
\label{remark1}
From (\ref{zipf2}), we notice that the value of the contents number affects the diversity of users' interest. It is hard to design a robust caching placement algorithm, which only depends on the statistical characteristics of content requests, to perform well in the varying contents number M. We tend to decide caching placement based on the real-time content request characteristics of networks.
\end{remark}

The users' requests may not be responded immediately and would be scheduled among time slots. In time slot $t$, ${{b_n}\left( t \right) = 1}$ if the requested contents by user $n$ is scheduled, otherwise ${{b_n}\left( t \right) = 0}$. The scheduled users at time slot $t$ will be transmitted for the requested content. Moreover, we assume that users will not request for new contents before their previous requests are responded. It means that we have ${{r_{nm}}\left( {t + 1} \right) = 1}$ if ${{r_{nm}}\left( t \right) - {b_n} > 0}$.

We assume the cache-enabling UAV could cache at most $Z$ contents and ${Z \le M}$. We assume the UAV proactively caches contents at each time slot.
If the requested content is cached by the UAV, it will be transmitted to users via radio access links directly, otherwise the content will be transmitted from the MBS to the UAV via wireless backhaul links. The UAV only need to fetch the content from the MBS once if it is not cached, even though there are several users requesting for the same content at the same time slot. Besides, contents may be transmitted from the MBS to the UAV because of proactive caching.
We define proactive caching index ${i_m}\left( t \right)=1$ to indicate content $m$ be cached at time slot $t$ for later time slots.
At the end of each time slot $t$, cache status is updated according to the proactive caching index of current time slot ${{i_m}\left( t \right)}$. If content $m$ has been cached in the UAV at the beginning of time slot ${t}$, ${{c_m}\left( t \right) = 1}$, otherwise ${{c_m}\left( t \right) = 0}$. Obviously, the proactive caching index at time slot $t$ decides the cached contents at time slot $t+1$, i.e. ${i_m}\left( t \right)={c_m}\left( t+1 \right)$.

\subsection{Channel Model}
We assume the downlink transmission consists of two parts, the wireless backhaul links from the ground MBS to the UAV, and radio access links from the UAV to the ground users.
As indicated in~\cite{R48675384}, the UAV-to-ground links can be modeled by a probabilistic path loss model.

Referring to the 3GPP specifications in~\cite{3GPP}, the path-loss of the wireless backhaul link is denoted as ${
{\bar g_{\rm ma}}\left( t \right)}$, which is randomly determined by line-of-sight (LoS) and non-line-of-sight (NLoS) link states
\begin{equation}
\label{3}
{\begin{array}{l}
{g_{\rm maLoS}} = 22.25{\log _{10}}{d_{\rm ma}}\left( t \right)\\
 - 0.5{\log _{10}}h{\log _{10}}{d_{\rm ma}}\left( t \right) + 20{\log _{10}}f + 30.9,
\end{array}}
\end{equation}
and
\begin{equation}
\label{4}
{\begin{array}{l}
{g_{\rm maNLoS}} = 43.2{\log _{10}}{d_{\rm ma}}\left( t \right)\\
 - 7.6{\log _{10}}h{\log _{10}}{d_{\rm ma}}\left( t \right) + 20{\log _{10}}f + 32.4,
\end{array}}
\end{equation}
where ${{d_{\rm ma}}\left( t \right)}$ represents the distance between MBS and UAV at time slot ${t}$, and ${f}$ represents the frequency of carrier. The probability of LoS is given in (5) at the bottom of this page, where ${{d_o} = \max \left[ {294.05{{\log }_{10}}h - 432.94,18} \right]}$, and ${{p_1} = 233.98{\log _{10}}h - 0.95}$. Then, the probability of NLoS could be calculated by ${\Pr \left( {{g_{\rm maNLoS}}\left( t \right)} \right) = 1 - \Pr \left( {{g_{\rm maLoS}}\left( t \right)} \right)}$.
Hence, the average path-loss can be expressed as
\setcounter{equation}{5}
\begin{equation}
\label{8}
{\begin{array}{l}
	{\overline g _{\rm ma}}\left( t \right) = \Pr \left( {{g_{\rm maLoS}}\left( t \right)} \right) \times {g_{\rm maLoS}}\left( t \right)\\
+ \Pr \left( {{g_{\rm maNLoS}}\left( t \right)} \right) \times \max \left\{ {{g_{\rm maLoS}}\left( t \right),{g_{\rm maNLoS}}\left( t \right)} \right\}.
	\end{array}}
\end{equation}

We assume the wireless backhaul links and the radio access links are allocated non-overlap frequency channels, and therefore have no co-frequency interference between them.
The signal-to-interference-plus-noise ratio (SINR) of the wireless backhaul link at time slot $t$ is
\begin{align}
{{\Gamma _{\rm B}}\left( t \right) = \frac{{{p_{\rm ma}}{{10}^{ - {{\bar g}_{\rm ma}}\left( t \right)/10}}}}{{{\sigma ^2} + \sum {{p_{\rm ma'}}{{10}^{ - {{\bar g}_{\rm ma'}}\left( t \right)/10}}} }}},
\end{align}
where ${{\sigma ^2}}$ represents the variance of additive Gaussian noise (AWGN), ${{p_{\rm ma}}}$ is the transmission power of the MBS, ${{{p_{\rm ma'}}}}$ is the transmission power of the neighboring MBSs in the networks, and  ${{{{\bar g}_{\rm ma'}}\left( t \right)}}$ is the path-loss from the neighboring MBS to the UAV.

\begin{figure} [t!]
	\centering
	\includegraphics[width= 3.5in]{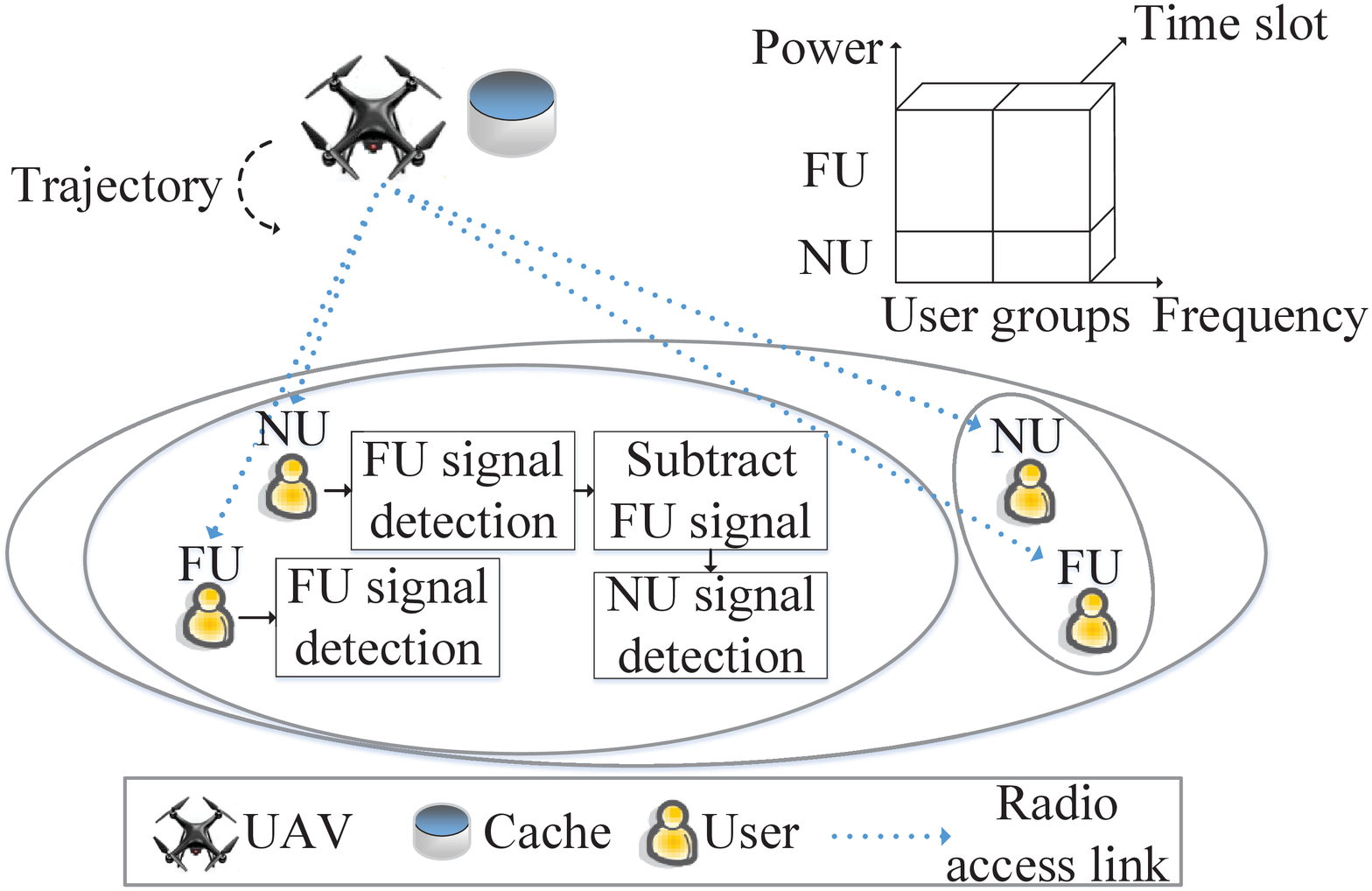}
	\caption{{NOMA based radio access links from UAV to ground users.}
	}
	\label{fig2noma}
\end{figure}

We can obtain the average path-loss between the UAV and the NU/FU of group $g$ at time slot $t$, ${\overline g _{{\rm NU}g}}\left( t \right)$ and ${\overline g _{{\rm FU}g}}\left( t \right)$, which are modeled following (\ref{8}).
The transmission power of the UAV is ${{p_{\rm uav}}}$, which is a constant during the flying duration, and is evenly allocated to user groups. The power of the UAV allocated to user group $g$ is expressed as ${{p_g}}$. At time slot $t$, the power allocation coefficient of the UAV to the NU in group $g$ is denoted as ${{h_g}\left( t \right)}$, which follows the proportion of power allocation. The power allocation coefficient of the UAV to the FU in group $g$ could be expressed as ${1 - {h_g}\left( t \right)}$.
The radio access links are based on NOMA, as shown in Fig.~\ref{fig2noma}.
The received signal at the NU is given by
\begin{align}
{\begin{array}{l}
{y_{{\rm NU}g}}\left( t \right) = \sqrt {{p_g}{h_g}\left( t \right)} {x_{{\rm NU}g}}\left( t \right){10^{{{ - {{\bar g}_{{\rm NU}g}}\left( t \right)} \mathord{\left/
 {\vphantom {{ - {{\bar g}_{{\rm NU}g}}\left( t \right)} {10}}} \right.
 \kern-\nulldelimiterspace} {10}}}} + \\
\sqrt {{p_g}\left( {1 - {h_g}\left( t \right)} \right)} {x_{{\rm FU}g}}\left( t \right){10^{{{ - {{\bar g}_{{\rm NU}g}}\left( t \right)} \mathord{\left/
 {\vphantom {{ - {{\bar g}_{{\rm NU}g}}\left( t \right)} {10}}} \right.
 \kern-\nulldelimiterspace} {10}}}} + {\zeta _{{\rm NU}g}}\left( t \right),
\end{array}}
\end{align}
where $\sqrt {{p_g}{h_g}\left( t \right)} {x_{NUg}}\left( t \right) + \sqrt {{p_g}\left( {1 - {h_g}\left( t \right)} \right)} {x_{FUg}}\left( t \right)$ represents the composite signal transmitted to the users in group $g$. ${\bar g_{{\rm NU}g}}\left( t \right)$ represents path loss of the radio access link to the NU of group $g$.

As shown in Fig.~\ref{fig2noma}, there exists interference between the NU and the FU in each group.
The NU desires to decode and remove the interference from the FU's superposition signal based on SIC. The interference cancellation is successful if the NU's received SINR for the FU's signal is larger or equal to the received SINR of the FU for its own signal~\cite{7460209,7812683}. We define the interference from superposition signal of the NU to the FU in receiver of the FU as ${{I_{\rm NF}} = {p_g}{h_g}\left( t \right){10^{ - {{\bar g}_{{\rm FU}g}}\left( t \right)/10}}}$.
The received SINR of the NU of group $g$ is
\begin{align}
\label{gaNU}
{{\Gamma _{{\rm NU}g}}(t) = \frac{{{p_g}{h_g}\left( t \right){{10}^{ - {{\bar g}_{{\rm NU}g}}\left( t \right)/10}}}}{{{\sigma ^2}}}}.
\end{align}
The SINR received at the FU of group ${g}$ is
\begin{align}
\label{gaFU}
{{\Gamma _{{\rm FU}g}}(t) = \frac{{{p_g}\left( {1 - {h_g}\left( t \right)} \right){{10}^{ - {{\bar g}_{{\rm FU}g}}\left( t \right)/10}}}}{{{I_{\rm NF}} + {\sigma ^2}}}}.
\end{align}

\subsection{Transmission Model}
In order to capture the dynamic of contents transmission and corresponding backhaul link delay from the MBS to the UAV, ${M}$ content virtual queues are defined to represent the contents waiting to be transmitted from the MBS to the UAV.
We define a backhaul transmission index of requested content $m$ as $\rho_{m}\left( t \right)$,
 which is calculated as ${\rho_{m}\left( t \right) = \left( {1 - {c_m}\left( t \right)} \right)\sum\nolimits_{n = 1}^N {{r_{nm}}\left( t \right)} }$.
It means ${{c_m}\left( t \right) = 1}$ if the requested content $m$ has been cached, then content $m$ would not be transmitted via the backhaul link, i.e., ${{\rho_{m}}\left( t \right) = 0}$. Otherwise, the UAV would fetch the content $m$ from the MBS if the content has not been cached, ${{\rho_{m}}\left( t \right) = 1}$.
Besides, when content $m$ is proactively cached at time slot $t$, the UAV will fetch the content from the MBS if the content has not been cached, which is represented as ${{i_{m}}\left( t \right)\left( {1 - {c_m}\left( t \right)} \right)}$. Hence, the virtual queue backlog for content ${m}$, denoted as ${{\mu _m}\left( t \right)}$, evolves over time slot as
\begin{equation}
\label{transmission}
{{\mu _m}\left( t \right) = \min \left[ {{\rho _m}\left( t \right) + {i_m}\left( t \right)\left( {1 - {c_m}\left( t \right)} \right),1} \right]},
\end{equation}
where ${\min \left[ * ,1\right]}$ means that for the same content ${m}$, the UAV only needs to request from the MBS once in a time slot.

The total bandwidth of the backhaul link is ${{B_{\rm B}}}$.  The backhaul transmission rate from the MBS to the UAV at time slot ${t}$ is
\begin{align}
{{R_{\rm B}}\left( t \right) = {B_{\rm B}}{\log _2}\left( {1 + {\Gamma _{\rm B}}\left( t \right)} \right)}.
\end{align}
As the bandwidth is equally allocated among content virtual queues, backhaul transmission delay of the requested content ${m}$ at time slot ${t}$ is expressed as
\begin{align}
{{D_{{\rm {\rm B}}m}}\left( t \right) = \left\{ {\begin{array}{*{20}{c}}
{0,if~\sum\nolimits_{m = 1}^M {{\mu _m}\left( t \right)}  = 0},\\
{\frac{{C_1}{{\mu _m}\left( t \right)}}{{\frac{1}{{\sum\nolimits_{m = 1}^M {{\mu _m}\left( t \right)} }}{R_{\rm B}}\left( t \right)}},if~\sum\nolimits_{m = 1}^M {{\mu _m}\left( t \right)}  \ne 0},
\end{array}} \right.}
\end{align}
where ${C_1}$ is the size of each content.


Consider the transmission of the radio access links from the UAV to the users at time slot ${t}$, as we define above, ${{b_n}\left( t \right) = 1}$ represents that the requested content of user $n$ is scheduled at time slot $t$. Thus, the frequency band of the radio access link will be fairly allocated to the NOMA user groups.  We define the total radio access  bandwidth of the UAV as ${{B_{ A}}}$. According to (\ref{gaNU}) and (\ref{gaFU}), the data rates of the NU in group $g$ is
\begin{align}
{{R_{{\rm NU}g}}\left( t \right) = \frac{{2{B_{\rm A}}}}{{\sum\nolimits_{n = 1}^N {{b_n}\left( t \right)} }}{\log _2}\left( {1 + {\Gamma _{{\rm NU}g}}\left( t \right)} \right)},
\end{align}
and that of the FU, ${{R_{{\rm FU}g}}\left( t \right)}$, can be calculated by the similar way.
The radio access transmission delay of user $n$ is
\begin{align}
{{D_{{\rm{A}}n}}\left( t \right) = {\varphi _{{\rm NU}g}}\left( t \right)\frac{C_1}{{{R_{{\rm NU}g}}\left( t \right)}} + {\varphi _{{\rm FU}g}}\left( t \right)\frac{C_1}{{{R_{{\rm FU}g}}\left( t \right)}}},
\end{align}
where ${{\varphi _{{\rm NU}g}}\left( t \right) = 1}$ if user ${n}$ corresponds to the NU of group ${g}$, otherwise ${{\varphi _{{\rm NU}g}}\left( t \right) = 0}$. Similarly, ${{\varphi _{{\rm FU}g}}\left( t \right) = 1}$ if the user ${n}$ corresponds to the FU in group ${g}$, otherwise ${{\varphi _{{\rm FU}g}}\left( t \right) = 0}$.

In this model, the users' requests may not be responded immediately and would be scheduled among time slots.
Therefore, the scheduling delay of content requesting users could be expressed as ${\left( {\sum\nolimits_{m = 1}^M {{r_{nm}}\left( t \right)}  - {b_n}\left( t \right)} \right)\delta}$, where ${\delta }$ is the length of time slot.
Considering the quality of experience (QoE) of users, we assume an upper limit of scheduling delay, denoted as ${\beta }$.

\subsection{Problem Formulation}
Given the above models, our goal is to minimize the long-term content delivery delay in dynamic networks. To achieve this goal, we formulate a problem by jointly optimizing the caching placement of the UAV, the user scheduling of content requests, and the power allocation of NOMA users.

The considered sum content delivery delay of users at each time slot consists of transmission delay of backhaul link, transmission delay of downlink radio access link, and scheduling delay of content requesting users, and therefore can be expressed as
\begin{align}
\label{costfunction}
{\begin{array}{l}
u\left( t \right) = \underbrace {\sum\limits_{m = 1}^M {{D_{{\rm B}m}}\left( t \right)} }_{backhaul~delay} + \underbrace {\sum\limits_{n = 1}^N {{D_{{\rm A}n}}\left( t \right)} }_{radio~access~delay} + \\
\underbrace {\sum\limits_{n = 1}^N {\left( {\sum\limits_{m = 1}^M {{r_{nm}}\left( t \right)}  - {b_n}\left( t \right)} \right)} \delta }_{scheduling~delay}
\end{array}},
\end{align}
where we ignore the uplink transmission delay and the processing delay.
Obviously, those three parts can be zero. For example, backhaul delay will be zero if the requested content is cached by the UAV, and scheduling delay will be zero if request of users is responded immediately.
\begin{remark}
\label{remark2}
From (\ref{costfunction}), we notice that the user number ${N}$ affects the content delivery delay of networks. The larger ${N}$ causes an increase in radio access delay and scheduling delay.
\end{remark}
According to (\ref{costfunction}), the long-term content delivery delay minimization problem can be expressed as
\begin{subequations}\label{optimal1}
	\begin{align}
	&\min_{b,{i},h}  {\kern 1pt}{\kern 1pt}{\kern 1pt}{\sum\limits_{t = 1}^T {u\left( t \right)} } \quad\\
	\label{25b}
	&{\kern 1pt}{\kern 1pt}{\kern 1pt}{\kern 1pt}{\kern 1pt}{\rm{s}}{\rm{.t}}{\rm{.}}{\kern 1pt} {\kern 1pt} {\kern 1pt} {\kern 1pt}{b_n}\left( t \right) \in \{ 0,1\} ,\forall n,t,\\
	\label{25c}
	&{\kern 1pt}{\kern 1pt}{\kern 1pt}{\kern 1pt}{\kern 1pt}{\kern 1pt}{\kern 1pt}{\kern 1pt}{\kern 1pt}{\kern 1pt}{\kern 1pt}{\kern 1pt}{\kern 1pt}{\kern 1pt}{\kern 1pt}{\kern 1pt}{\kern 1pt}{\kern 1pt}{\kern 1pt}{\kern 1pt}{\kern 1pt}{\kern 1pt}{\kern 1pt}{i_m}\left( t \right) \in \{ 0,1\} ,\forall m,t,\\
    \label{25d}
	&{\kern 1pt}{\kern 1pt}{\kern 1pt}{\kern 1pt}{\kern 1pt}{\kern 1pt}{\kern 1pt}{\kern 1pt}{\kern 1pt}{\kern 1pt}{\kern 1pt}{\kern 1pt}{\kern 1pt}{\kern 1pt}{\kern 1pt}{\kern 1pt}{\kern 1pt}{\kern 1pt}{\kern 1pt}{\kern 1pt}{\kern 1pt}{\kern 1pt}{\kern 1pt}0 \le {h_{g}}\left( t \right) \le 1,\forall t,g,\\
	\label{25e}
	&{\kern 1pt}{\kern 1pt}{\kern 1pt}{\kern 1pt}{\kern 1pt}{\kern 1pt}{\kern 1pt}{\kern 1pt}{\kern 1pt}{\kern 1pt}{\kern 1pt}{\kern 1pt}{\kern 1pt}{\kern 1pt}{\kern 1pt}{\kern 1pt}{\kern 1pt}{\kern 1pt}{\kern 1pt}{\kern 1pt}{\kern 1pt}{\kern 1pt}{\kern 1pt}{\kern 1pt}{{b_n}\left( t \right) \le \sum\limits_{m = 1}^M {{r_{nm}}\left( t \right)},\forall t,}\\
	\label{25f}
	&{\kern 1pt}{\kern 1pt}{\kern 1pt}{\kern 1pt}{\kern 1pt}{\kern 1pt}{\kern 1pt}{\kern 1pt}{\kern 1pt}{\kern 1pt}{\kern 1pt}{\kern 1pt}{\kern 1pt}{\kern 1pt}{\kern 1pt}{\kern 1pt}{\kern 1pt}{\kern 1pt}{\kern 1pt}{\sum\limits_{t - \beta  + 1}^t {\left( {\sum\limits_{m = 1}^M {{r_{nm}}\left( t \right)}  - {b_n}\left( t \right)} \right)}  < \beta  ,\forall n,t,} \\
	\label{25g}
	&{\kern 1pt}{\kern 1pt}{\kern 1pt}{\kern 1pt}{\kern 1pt}{\kern 1pt}{\kern 1pt}{\kern 1pt}{\kern 1pt}{\kern 1pt}{\kern 1pt}{\kern 1pt}{\kern 1pt}{\kern 1pt}{\kern 1pt}{\kern 1pt}{\kern 1pt}{\kern 1pt}{\kern 1pt}{\kern 1pt}{\kern 1pt}{\sum\limits_{m = 1}^M {{i_m}\left( t \right)}  \le Z,\forall t,}
	\end{align}
\end{subequations}
where constraints (\ref{25b}) and (\ref{25c}) show that the values of ${{b_n}\left( t \right)}$ and ${{i_m}\left( t \right)}$ should be either 0 or 1. Constraint (\ref{25d}) shows the range of power allocation coefficient ${{h_{g}}\left( t \right)}$ for the NOMA. Constraint (\ref{25e}) guarantees that only the users waiting for response could be responded. Considering QoE, constraint (\ref{25f}) is assigned to limit the scheduling delay. Constraint (\ref{25g}) guarantees the sum of the proactive cache scheme at time slot ${t}$ should be no more than cache capacity of UAV. It is obvious that formulated problem (\ref{optimal1}) is NP-hard, which is demonstrated in appendix \ref{appenA}.

\section{Reinforcement Learning Based Algorithm}
We convert problem (\ref{optimal1}) into a MDP to cope with the dynamic UAV locations and the content requests. Machine learning (ML) has emerged as a powerful artificial intelligence (AI) technique to make the UAV wireless communication highly efficient~\cite{R48926369}. Then the Q-learning based caching placement and resource allocation algorithm is used to tackle the MDP.
However, the efficiency of the Q-learning based algorithm is limited in the scenario with huge state/action space. Therefore, we use the function approximation based caching placement and resource allocation algorithm to solve the proposed problem for large-scale networks.
\subsection{Problem Conversion}
As the optimize objective of formulated problem is to minimize the content delivery delay in dynamic networks, we convert the formulated problem (\ref{optimal1}) into a MDP, where the UAV acts as an agent. The proposed MDP consists of four components: a) set of finite states, b) set of finite actions, c) dynamic change of states, which describes how current state and action influence the future state, and d) the cost function defined in (\ref{costfunction}).

Since we consider the networks are dynamic with the requests of users for $M$ contents in cache-enabling UAV networks,
the states of the MDP is
characterized by users waiting for response ${W\left( t \right)}$ and cache situation of UAV ${C\left( t \right)}$ as follows. ${W\left( t \right) = \left[ {{w_1}\left( t \right), \cdots ,{w_n}\left( t \right), \cdots ,{w_N}\left( t \right)} \right]}$ denotes the users waiting for response at time slot $t$, which consists of postponed users of previous time slots and users request for contents at time slot $t$. As mentioned above, ${{r_{nm}}\left( t \right)}$ is defined to represent request of user ${n}$ to content ${m}$ at time slot $t$. Thus, ${{w_n}\left( t \right)}$ could be calculated as ${{w_n}\left( t \right) = \sum\nolimits_{m = 1}^M {{r_{nm}}\left( t \right)}}$. ${C\left( t \right) = \left[ {{c_1}\left( t \right), \cdots ,{c_m}\left( t \right), \cdots ,{c_M}\left( t \right)} \right]}$ denotes the cache situation of $M$ contents at time slot $t$.
The state vector of the proposed MDP at time slot $t$ is defined as ${s\left( t \right) = \left[ {C\left( t \right),W\left( t \right)} \right]}$. The state space $S$ is equivalent to all possible combination of cache situation and users waiting for response.
The content request model can be extended to time varying user preference occasion, where the preference of users can be included in state variables to cope with the complex dynamic environment.

For the action of the MDP, we denote ${I\left( t \right) = \left[ {{i_1}\left( t \right), \cdots ,{i_m}\left( t \right), \cdots ,{i_M}\left( t \right)} \right]}$ as the proactive caching index of $M$ content at time slot $t$.
${B\left( t \right) = \left[ {{b_1}\left( t \right), \cdots ,{b_n}\left( t \right), \cdots ,{b_N}\left( t \right)} \right]}$ represents the scheduling to $N$ users at time slot $t$.
${{h_{g}}\left( t \right)}$ is the power allocation of NU of group $g$ at time slot $t$. In order to form a limited action space, we describe the power allocation coefficient of NU with ${O}$ discrete power levels, i.e., ${{h_{g}}\left( t \right) \in \left\{ {h_{}^1, \cdots ,h_{}^O} \right\}}$. The power allocation coefficient of FU could be calculated as ${1 - {h_{g}}\left( t \right)}$. Then power allocation of NOMA users could be represented by power allocation coefficient of NU. The discrete power levels are reasonable since the transmission power is discrete in practical communication equipments.
The action vector executed by UAV at time slot $t$ could be expressed as ${a\left( t \right) = \left( {I\left( t \right),B\left( t \right),{H}\left( t \right)} \right)}$, which consists of caching placement of UAVs, user scheduling of content requests, as well as power allocation of NOMA users at time slot ${t}$.
 We define action space as a set of all possible combinations of these three factors, which could be express as ${A = I \otimes B \otimes H}$, with ${\otimes }$ denoting the Cartesian product.

Then, the characteristic of dynamic state is introduced. At time slot ${t + 1}$, the environment would switch to a new state ${s\left( {t + 1} \right)}$, which is determined by previous state ${s\left( t \right)}$ and action selection ${a\left( t \right)}$.  First, cache condition of content $m$ at time slot $t+1$ could be updated according to caching placement of time slot ${t}$, and acts as the element of ${C\left( t+1 \right)}$. Then, users waiting for response could be calculated ${{w_n}\left( t+1 \right) = \sum\nolimits_{m = 1}^M {{r_{nm}}\left( t+1 \right)}}$. For convenience, rule of state switch is summarized as ${\xi \left( {s\left( t \right),a\left( t \right),s\left( {t + 1} \right)} \right)}$.
When the UAV takes an action ${{a\left( t \right)}}$ at time slot ${t}$, we could get the instantaneous cost considering current state ${s\left( t \right)}$. The instantaneous cost is defined as the instantaneous sum delay  ${u\left( t \right)}$, which is calculated according to (\ref{costfunction}).

The MDP could be described as tuple ${\Gamma  = \left( {S,A,\xi ,u} \right)}$.
The optimization objective of formulated problem is to minimize the long-term sum delay, which is solved by Q-learning.

\subsection{Q-Learning}
Q-learning is a RL method for solving the problem modeled after MDPs, where a learning agent operates in an unknown environment~\cite{8736350}. At each time slot $t$, the UAV acts as an agent to confirm current state ${s\left( t \right) \in S}$, select action ${a\left( t \right) \in A}$, and get a cost ${u\left( t \right)}$, according to the current state and the action selection.
Then, the environment switches to the next state ${s\left( {t + 1} \right)}$, according to the current state and the selected action.

Two fundamental concepts of the algorithm for solving the above MDP are state-value function and action-value function (Q-function)~\cite{ZhaoN19TWC}.
The state-value function $V$ is defined to measure the importance of states, which is ${V\left( {s,\pi } \right) = E\left\{ {\sum\limits_{\tau  = t}^{ + \infty } {{\gamma ^{\tau  - t}}u\left( \tau  \right)|s\left( t \right) = s} } \right\}}$,
where ${0 \le \gamma  < 1}$ denotes the discount factor.
The action-value function~(Q-function) is defined to measure the importance of action, which can be expressed as ${Q\left( {s,a,\pi } \right) = E\left\{ {\sum\limits_{\tau  = t}^{ + \infty } {{\gamma ^{\tau  - t}}u\left( \tau  \right)|s\left( t \right) = s,a\left( t \right) = a} } \right\}}$.
We can get the relationship between the state-value function and the action-value function~\cite{tesauro2004extending}
\begin{align}
{V\left( {s,\pi } \right) = \sum\limits_{a \in A} {\pi \left( {s,a} \right)Q\left( {s,a,\pi } \right)}},\label{39}
\end{align}
where $\pi \left( {a|s} \right)$ represents the probability that the agent with state $s$ select action $a$ at time slot $t$, i.e. ${\pi \left( {a|s} \right) = P\left( {{a_t} = a|{s_t} = s} \right)}$~\cite{8672604}.

As we introduced before, the target of the proposed MDP is to minimize the long-term cost by selecting the most suitable strategy. The optimal problem can be written as
\begin{align}
{{\pi ^*} = \mathop {\arg \min }\limits_\pi  V(s)},\label{40}
\end{align}
where we define ${\pi ^*}$ to express the optimal action selection scheme.
Combining (\ref{39}) and (\ref{40}), the optimal problem can be reformulated as
\begin{align}
\label{35}
{{V^*}(s) = \mathop {\min }\limits_a {Q^*}(s,a)}.
\end{align}
In other words, we can get the optimal state-value by selecting the action with the least Q-value.


The update rule of the action value function~\cite{singh1993convergence} can be denoted as
\begin{align}
\label{Qbell}
{\begin{array}{l}
{Q_{t + 1}}\left( {s,a} \right) = \left( {1 - \alpha \left( t \right)} \right){Q_t}\left( {s,a} \right) + \\
\alpha \left( t \right)\left\{ {u\left( t \right) + \gamma \mathop {\min }\limits_{a \in A} {Q_t}\left( {s',a'} \right)} \right\},
\end{array}}
\end{align}
where ${s}$ and ${a}$ represent the state and the action of time slot ${t}$ respectively, and ${s'}$ and ${a'}$ represent the state and the action of time slot ${t + 1}$ respectively.
%
%
%
%
%
%
%
%
According to~\cite{singh1993convergence}, we represent the learning rate as
\begin{align}
{\alpha \left( t \right) = \frac{1}{{{{(t + {c_\alpha })}^{{\varphi _\alpha }}}}}},
\end{align}
where ${{c_\alpha } > 0}$ and ${{\varphi _\alpha } \in ({\rm{1/2}},1]}$. The learning rate represents the impact of learning result to the Q-table.

Action selection mechanism is an important part of Q-learning. In this paper, we use the \emph{soft ${\varepsilon}$-greedy} method  to guide the action selection of the UAV. In the proposed method, instead of keeping a fixed probability of exploration in the action space, we define a gradually decreasing probability to randomly select actions,
\begin{align}
\label {31}
{\left\{ {\begin{array}{*{20}{c}}
		{1,{\rm{ }}~if~{\rm{  }}t \le \varepsilon ,}\\
		{\frac{\varepsilon }{t},{\rm{ }}~if~{\rm{  }}t > \varepsilon ,}
		\end{array}} \right.}
\end{align}
where ${\varepsilon }$ is a positive number. According to (\ref{31}), for the time slot smaller than ${\varepsilon }$, the UAV selects actions randomly. Besides, for the time slot bigger than ${\varepsilon }$, the UAV explores action space with a decreasing probability.
Larger ${\varepsilon }$ means that there would be more time slots to select an action randomly at the beginning of the iterations, which leads to a fast exploration of action space. When the action space has been explored in some extent, the Q-table would be more stable with a small probability of randomly select actions, as the random selection would influence the Q-value of actions of former time slots.

\begin{remark}
\label{remark3}
From (\ref{31}), we notice that the value of ${\varepsilon }$ in the Q-learning affects the trade-off between exploration of action space and exploitation of the explored result.
Larger ${\varepsilon }$ achieves a sufficient exploration, but the high proportion of randomly selecting action affects the stability of long-term Q value calculation.
Smaller ${\varepsilon }$ achieves a rapidly decreasing content delivery delay, beacause of the generous exploitation for the explored result. However, the insufficient exploration for the action space limits the converged content delivery delay.
We should point that the objective of this paper is not to investigate the optimal trade-off between exploration and exploitation by ${\varepsilon }$, which jointly decide convergence speed and converged value of content delivery delay.
\end{remark}

Moreover, legal action is defined in this paper. We could confirm that only a subset of the action set could be accessed for the specific states, due to the restrictions of the proposed problem. The actions that the agent could choose when the environment is in state ${s\left( t \right)}$ is defined as the legal action of that state.
Instead of listing all possible states and actions, state list and action list are initially empty, and then gradually increase. When a new state $s$ is experienced by the UAV, it will be saved in the state list. Besides, when a new action $a$ is selected by the UAV, it will be saved in the action list. As mentioned above, the legitimacy of the action varies with different states, so the legitimacy is judged before the action is taken. In particular, the actions in the list are judged in the descend order of Q-value until a legal action is found. The UAV will randomly take a legal action if there is no legal action in the list.

Based on Q-learning, iterative caching placement and resource allocation algorithm is summarized in Algorithm 1.
\begin{algorithm}
\caption{Q-learning based caching placement and resource allocation algorithm}
\label{alg1}

\begin{algorithmic}[1] 
\REQUIRE
\STATE Set parameters
\STATE Initialize state, Q-table, action list and state list.
\ENSURE
\WHILE{$t < T$}
\STATE Update state list.
\IF {{$t \le \varepsilon$ } or {$random[0,1] \le \varepsilon /t$}}
\STATE Select a legal action randomly.
\ELSE
\STATE Sort the listed actions according to Q-table.
\IF {the action with current optimal Q-value is legal}
\STATE Select the action.
\ELSE
\STATE Judge the action with sub-optimal Q-value and back to step (9).
\IF {the action with sub-optimal Q-value is None}
\STATE Select a legal action randomly.
\ENDIF
\ENDIF
\ENDIF
\STATE Update action list, cost, Q-table, state.
\ENDWHILE
\end{algorithmic}
\end{algorithm}

\subsection{Function Approximation Based Algorithm for Large-Scale Networks}
\begin{figure} [t!]
	\centering
	\includegraphics[width= 3.5in]{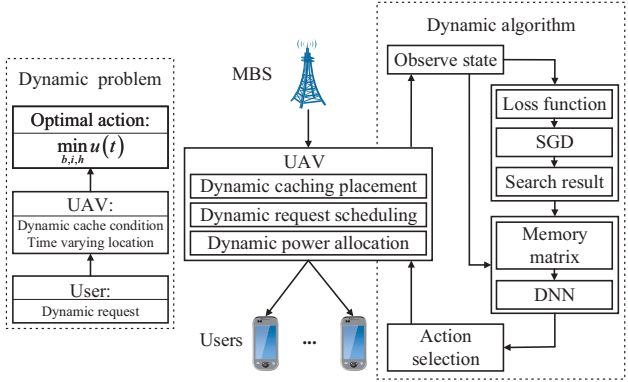}
	\caption{{Framework of the function approximation based algorithm.}
	}
	\label{s2fig3}
\end{figure}

In the proposed Q-learning based caching placement and resource allocation algorithm, the Q-table is too large to search and save in the large-scale networks since the size of the action space and the state space are all mainly related to the number of contents and users.
In this subsection, we use a function approximation based caching placement and resource allocation algorithm to solve the proposed problem for large-scale networks in practical scenarios.
The framework of the function approximation based algorithm is given in Fig.~\ref{s2fig3}, where the stochastic gradient descent method (SGD) is used to search the action space efficiently and the deep neural networks (DNN) is adopted to overcome the limitations of Q-table storage.

In the traditional reinforcement learning algorithm, ${\varepsilon}$-greedy is deployed as an action selection scheme~\cite{tesauro2004extending,Cui2018Multi}, which is not efficient enough to search the large-scale action space in the formulated MDP of this paper. The insufficient search result will lead to a limitation for DNN training. Thus the large action space in our problem is searched with SGD, which is an efficient algorithm with low complexity.

In the proposed algorithm, we use the SGD to search the action space, whose result is stored in the memory matrix. The mappings between states and actions stored in the matrix act as the supervisors for training of the DNN.
Loss function is defined as
\begin{align}
{\begin{array}{l}
u\left( {I,B,H} \right) = \sum\limits_{m = 1}^M {\frac{{{w_m}\left( {{b_n},{i_m}} \right)}}{{{R_{\rm B}}\left( {{b_n},{i_m}} \right)}} + \sum\limits_{n = 1}^N {\left( {\sum\limits_{m = 1}^M {{r_{nm}}}  - {b_n}} \right)} \delta  + } \\
\sum\limits_{n = 1}^N {\left( {{\varphi _{{\rm NU}g}}\frac{{{C_1}}}{{{R_{{\rm NU}g}}\left( {{b_n},{h_g}} \right)}} + {\varphi _{{\rm FU}g}}\frac{{{C_1}}}{{{R_{{\rm FU}g}}\left( {{b_n},{h_g}} \right)}}} \right)},
\end{array}}
\end{align}
where ${I = \left\{ {{i_1}, \cdots ,{i_M}} \right\}}$, ${B = \left\{ {{b_1}, \cdots ,{b_N}} \right\}}$, and ${H = \left\{ {{h_1}, \cdots ,{h_G}} \right\}}$. Besides, discrete variables in vector ${\left\{ {I,B,H} \right\}}$ are relaxed to continuous variables.
Variables ${{i_m}}$, ${{b_m}}$, and ${{h_g}}$ are updated according to the SGD method in~\cite{8272511} to minimize the loss function.

There are constraints to solution ${\left\{ {I,B,H} \right\}}$ of the optimization problem. Before storing the mappings in the memorize matrix, we transform the output continuous values to the form that meets the constraints.
According to (\ref{25c}) and (\ref{25g}), we sort all the elements of ${I = \left\{ {{i_1}, \cdots ,{i_M}} \right\}}$ in the descending order and the ${j_2}$th element is ${{i_{{s_{j_2}}}}}$, which corresponds to the ${{s_{j_2}}}$th content before sorting for ${{j_2} = 1,2, \cdots ,M}$. Then, the largest $Z$ elements are set as 1, i.e. ${{i_{{s_{j_2}}}} = 1}$ for ${{j_2} = 1, \cdots ,Z}$, where $Z$ is the cache capacity of the UAV. Besides, we convert ${{b_n}}$ to Boolean variable with $0.5$ as the borderline, according to (\ref{25b}). According to (\ref{25d}), continuous variables ${H}$ are discretized. Then, mapping between the current state and action selection is stored in the memory matrix, where the capacity of the matrix is limited. In particular, if the memory matrix is filled, the longest remembered mapping will be replaced by the mapping of time slot ${t}$.

In the traditional reinforcement learning algorithm, Q-table is a common method to store the mappings between states and actions. The large state space and action space in the formulated MDP of this paper make the Q-table difficult to store and search. Thus we use a DNN to approximate the relationship between states and actions.

The action selection is treated as a black box and the DNN is deployed to learn relation between states and actions because of DNN's universal approximation ability~\cite{deeplearning}. The proposed algorithm contains two stages, training process and testing process. The formulated DNN is trained by memory replay. During the training process of the DNN based function approximation model, samples in the memory matrix act as the supervisor. During the testing process, we could select the optimal action for state $s$ according to the output of the model.

In the DNN based function approximation model, the mappings in memory matrix are replied to train the neural network, where the state and action samples act as the supervisor. Moreover, optimal action selection could be decided by DNN, with state vector $s$ as input.
The agent in our DNN is the UAV. The DNN uses an input vector ${X\left( t \right) = \left[ {C\left( t \right),W\left( t \right)} \right]}$ to represent the state at time slot ${t}$.
Moreover, the output vector of the DNN is ${Y\left( t \right) = \left[ {{y_1}\left( t \right), \cdots ,{y_K}\left( t \right)} \right]}$, which represents the user scheduling of content requests, caching placement of UAV, as well as power allocation of NOMA users for the current state, and ${K = M + N + N/2}$. We respectively set the input layer and the output layer as ${Ne{t_X} \in {R^{\left( {M + N} \right) \times 1}}}$, and ${Ne{t_Y} \in {R^{K \times 1}}}$. The hidden layers consist of feature maps ${Ne{t_a} \in {{\mathbb{R}}^{{4} \times {4}}}}$, ${Ne{t_b} \in {{\mathbb{R}}^{{2} \times {2}}}}$ and fully connected layer ${Ne{t_c} \in {{\mathbb{R}}^{{4} \times 1}}}$.

\begin{table*}[!t]
	\small
	\centering
	\caption{Time complexity and execution time of algorithms}
	\label{tabelcomplexity}
	\begin{tabular}{|c|c|c|}
		\hline
		Algorithm                                & Time complexity                                                                                                    & Execution time \\ \hline
		Q-learning based algorithm               & $O\left( {\sum\nolimits_{{\tau _1} = 1}^{{T_1}} {{A_e}\left( {M,N,{Z},{\tau _1}} \right)} } \right)$             & 593.774 s      \\ \hline
		Function approximation based algorithm   & $O\left( {{T_1}\max \left( {{T_2}\left( {1.5N + M} \right),{T_3}{{\left( {1.5N + M} \right)}^2}} \right)} \right)$ & 312.727 s      \\ \hline
		Greedy based exhaustive search algorithm & $O\left( {{T_1}A\left( {M,N,{Z}} \right)\max \left( {M,N} \right)} \right)$                                      & 104482.649 s   \\ \hline
		Fixed algorithm                          & $O\left( {{T_1}\max \left( {M,N} \right)} \right)$                                                                 & 99.313 s       \\ \hline
		Random algorithm                         & $O\left( {{T_1}\max \left( {M,N} \right)} \right)$                                                                 & 104.915 s      \\ \hline
	\end{tabular}
\end{table*}

The DNN model consists of the input weight matrix ${{W_a} \in {{\mathbb{R}}^{\left( {M + N} \right) \times 16}}}$,
convolution kernel matrix ${{W_b} \in {{\mathbb{R}}^{3 \times 3}}}$, weight matrix
${{W_c}}$, and ${{W_d}}$.
Besides, activation function can be rectified linear unit, ${{f_{relu}}\left( x \right) = \max \left( {0,x} \right)}$, and batch normalization is deployed to maintain the network stability. We employ gradient truncation to prevent gradient explosion. The bias matrices have the same size of the corresponding weight matrices.

In order to build the relationship between the input ${Ne{t_X}}$ and the output ${Ne{t_Y}}$, weight matrices and  bias matrices need to be trained.
During the traning stage, the model is trained to minimize the distance between the DNN's output and the action in memory, which can be measured by
\begin{align}
{{L_1}\left( t \right) = \frac{1}{K}\sum\limits_{k = 1}^K {{{\left( {{y_k}\left( t \right) - {y_k}\left( t \right)'} \right)}^2}} },
\end{align}
where ${Y' = \left[ {{y_1}', \cdots ,{y_k}'} \right]}$ represents the action in the memory matrix. Then weight matrices and bias matrices could be respectively updated according to SGD.
When the agent is faced with a state, the proposed DNN model can output a vector, which contains user scheduling of content requests, caching placement of UAV, as well as power allocation of NOMA users.

\begin{algorithm}
	\caption{Function approximation based caching placement and resource allocation algorithm}
	\label{alg2}

\begin{algorithmic}[1] 
	\REQUIRE
	\STATE Randomly initialize starting state including cache situation ${W\left( t \right)}$, and users waiting for response ${C\left( t \right)}$. Set time slot ${t=1}$, The period for the reset of DNN is $T$;
	\ENSURE
	\FOR{time slot $t$}
	\STATE  Select an action randomly.
	\FOR{iteration ${\tau _{\rm{1}}} $}
	\STATE \textbf{SGD}: Update the action according to SGD.
	\ENDFOR
	\STATE Store current state and optimal action selection according to SGD in the memory matrix.
	\IF{$t$ is a multiple of $T$}
	\STATE \textbf{DNN Reset}: Randomly initialize networks nodes ${Ne{t_X}}$, ${Ne{t_a}}$, ${Ne{t_b}}$, ${Ne{t_c}}$, ${Ne{t_Y}}$, weight matrices ${{W_a}}$, ${{W_b}}$, ${{W_c}}$, ${{W_d}}$, and bias matrices.
	\FOR{iteration
${\tau _{\rm{2}}}$}
	\STATE \textbf{DNN Training}: Update ${Ne{t_X}}$, ${Ne{t_a}}$, ${Ne{t_b}}$, ${Ne{t_c}}$, ${Ne{t_Y}}$, ${{W_a}}$, ${{W_b}}$, ${{W_c}}$, ${{W_d}}$, and bias matrices according to the mechanism introduced above.
	\ENDFOR
	\ENDIF
	\STATE \textbf{Make decision}: Recursively calculate ${Ne{t_Y}}$. Convert the value of $Ne{t_Y}\left( {1:M + N,1} \right)$ to a Boolean variable, and convert the value of $Ne{t_Y}\left( {M + N + 1:M + N + N/2,1} \right)$ to discrete values.
    \ENDFOR
\end{algorithmic}
\end{algorithm}

Based on the function approximation method, the proposed dynamic caching placement and resource allocation algorithm is summarized in Algorithm 2.
\subsection{Analysis of the Proposed Algorithms}
{\bfseries 1. Complexity :} The time complexity of the Q-learning based algorithm, function approximation based algorithm, greedy based exhaustive search algorithm, fixed algorithm, and random algorithm are listed in Table \ref{tabelcomplexity}. The number of considered time slot is $T_1$, we defined ${{A_e}\left( {M,N,{Z},{\tau _1}} \right)}$ to express the size of the explored action space, which is jointly determined by the size of the complete action space and current time slot ${\tau _1}$. We define $A\left( {M,N,{Z}} \right)$ to express the size of the complete action space, which is jointly determined by the number of users, the number of contents and the cache capacity of UAV. Considering the function approximation based algorithm, ${T_2}$ in ${T_2}\left( {1.5N + M} \right)$ represents the converged iteration of the SGD algorithm, ${{1.5N + M}}$ represents the dimension of the variables to optimize. ${T_3}$ in ${T_3}{\left( {1.5N + M} \right)^2}$ represents the size of the mini batch for DNN training, where ${\left( {1.5N + M} \right)^2}$ represents the largest dimension of DNN weight matrices. We define $\max \left( {} \right)$ to express that the bigger complexity is selected for the calculation of the total time complexity, because of the cascade relationship between SGD algorithm and the training of the DNN. Considering the greedy based exhaustive search algorithm, we define $A\left( {M,N,{Z}} \right)$ to express the size of the complete action space, which is determined by the number of users $N$, the number of contents $M$, and the cache capacity of UAV $Z$. As the cost of every action is calculated in the greedy based exhaustive search algorithm, the complexity of every calculation is defined as ${\max (M,N)}$. We also compare the execution time of algorithms with iteration ${T_1} = {10^5}$ in the matlab simulation software, where $M=8$, $N=4$, $Z=2$, $T_2=200$ and $T_3=32$. The type of central processing unit of the laptop is Intel(R) Core(TM) i5-7300HQ, with the calculation frequency of 2.50 GHz. The length of time slot is 0.05s, which is assumed according to the movement of the UAV. However, the execution time of the greedy based exhaustive search algorithm is much bigger than the sum length of time slots. This is because that the exhaustive search algorithm puts forward high requirement on the performance of the computer, which proves the importance of proposing algorithms to get the trade-off between performance and complexity.

{\bfseries 2. Compatibility of the proposed solution to the networks with multi-MBSs :} The solution proposed in this paper is also suitable for the networks with multi-MBSs. For the multi-cell networks with orthogonal frequency resource, the solution proposed in this paper can be applied directly. For the multi-cell networks with  frequency reuse, the SINR of transmission links is influenced by the interference from neighbor cells' MBSs and UAVs. The optimization problem for the cellular networks with multiple MBSs and UAVs is relatively complicated, which can be splitted into multiple optimization problems focused on resource allocation optimization of a single UAV [44]. The proposed algorithm and the formulated problem can be applied to the cellular networks with multi-MBSs.

\section{Performance Evaluation}
The simulation results are provided in this section to verify the proposed algorithms. We consider a cellular network with multiple MBSs and multiple users. One MBS with heavy traffic load is selected as the target MBS, which is aided by a mobile UAV for traffic offloading and content caching.

In the simulation, we consider one target cell with six neighbor cells, where the side length is 100 m. We assume that the MBSs are loacted in the central of each cell. Users are ramdomly distributed in the cell of the target MBS. During the considered period, the users' location remains the same.
We assume that the flight trajectory of the UAV is predetermined in a circle [29] with 200 $m$ diameter and random central point in the cell of the target MBS. We assume that the speed of the UAV is 20m/s. The length of time slot considered in this paper is $\delta  = 0.05 s$.
The  upper limit of scheduling slot is assumed to be ${\beta  = 2}$.
We assume the request generate coefficient ${{R_g} = 2}$, which can be extended to various request generate coefficients with the limitation of ${{R_g} \le \frac{N}{\beta }}$.
The scheduling method for the users with requests in the same time slot is based on frequency division multiplexing.
There is a content library with $M$ unified contents with the same data size. The size of each unified content is $C_1=2$ MB. The cache space of the UAV is $Z=2$. The system can be extended to various content size case easily.
The different content files with various data size can be reshaped into unified contents by UAV caching.
The number of power allocation levels to NOMA users is $5$, which is expressed as ${{h_{g}}\left( t \right) \in \left\{ {0.1,0.2,0.3,0.4,0.5} \right\}}$.
The main simulation parameters follow the 3GPP specifications~\cite{3GPP}, which are summarized in Table \ref{tabel2}.
\begin{table}[!t]
	\small
	\caption{System Parameters}
\label{tabel2}
	\begin{center}
		\begin{tabular}{|l|l|}
			\hline
			Power of MBS ${{p_{\rm ma}}}$               & ${46}$ dBm    \\
			\hline
			Power of UAV ${{p_{\rm uav}}}$                         & $30$ dBm       \\ \hline
			Bandwidth of backhaul link ${{B_{ B}}}$            & $20$ MHz     \\ \hline
			Bandwidth of radio access link ${{B_{ A}}}$            & $20$ MHz     \\ \hline
			Noise power ${\sigma }$                            & $-174$ dBm/Hz  \\ \hline
			UAV flight altitude  ${h}$                       & $100$ m        \\ \hline
			Long-term period ${T}$                            & $100$
			\\ \hline
		\end{tabular}
	\end{center}
\end{table}

We compare the proposed Q-learning (QL) based algorithm and function approximation (FA) based algorithm with the benchmark algorithms to evaluate their effectiveness.
The benchmark algorithms are defined as follows.
\begin{itemize}
	\item Greedy algorithm: the UAV selects the optimal action in current state by exhaustive search to get the optimal instantaneous content delivery delay of current state;
	\item Fixed algorithm: the UAV caches the most popular contents in previous states, schedules the requesting users with round robin method, and allocates fixed power level for NOMA users;
	\item Random algorithm: the UAV selects actions randomly for content caching and radio resource allocation.
\end{itemize}

\begin{figure} [t!]
	\centering
	\includegraphics[width= 3.5in]{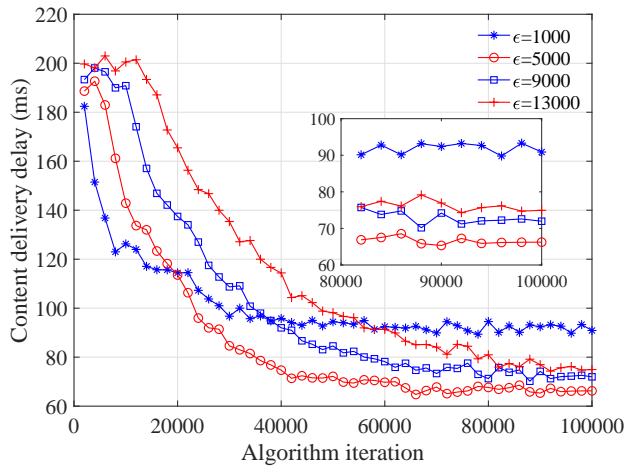}
	\caption{{Convergence of content delivery delay.}
	}
	\label{fig4re1}
\end{figure}

\begin{figure} [t!]
	\centering
	\includegraphics[width= 3.5in]{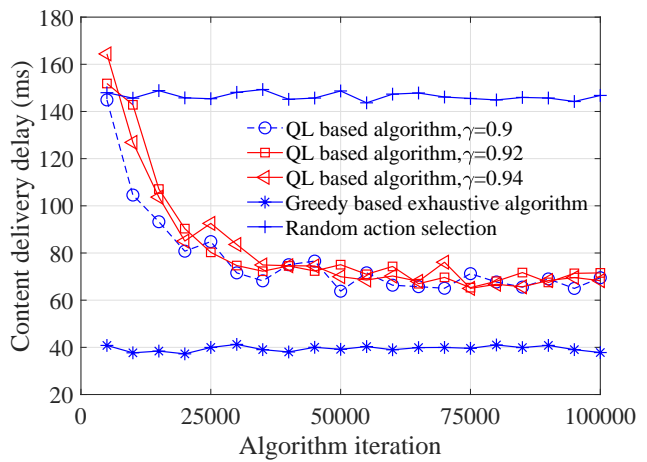}
	\caption{{Content delivery delay comparison with iteration numbers.}
	}
	\label{fig5re2}
\end{figure}

First, we verify the convergence of the proposed QL based algorithm by Fig.~\ref{fig4re1} and Fig.~\ref{fig5re2}. In this simulation, we set ${M = 4}$ and ${N=8}$.
Fig.~\ref{fig4re1} depicts the content delivery delay convergence of the proposed QL based algorithm with different action selection parameters ${\varepsilon }$ with the discount factor ${\gamma}$ as 0.9. In the proposed algorithm, the agent selects the action with the smallest Q-value in addition to exploration. 
In Fig.~\ref{fig4re1}, the content delivery delay converges gradually with the number of iterations increasing, ignoring the shaking caused by the state transition.
Moreover, Fig.~\ref{fig4re1} indicates that the content delivery delay with different ${\varepsilon}$ has different convergence speed and converged value.
It is observed that content delivery delay of the algorithm with ${{\varepsilon} {\rm{ = 13000}}}$ decreases slowly at the beginning of iterations.
This is because that althought the action space is generously explored, but content delivery delay decreases slowly with insufficient exploitation of the explored actions. Because of the high proportion of selecting action randomly, the Q value calculating of the algorithm with ${{\varepsilon} {\rm{ = 13000}}}$ is unstable. As a result, the converged content delivery delay of the algorithm is relatively bigger.
As we observe from Fig.~\ref{fig4re1}, the algorithm with smaller ${{\varepsilon} {\rm{ = 1000}}}$ achieves a rapid decrease of content delivery delay at the beginning of iterations, as the exploration result is efficiently exploited. However, converged content delivery delay of the algorithm is relatively bigger because of the insufficient exploration of the action space.
In the simulation, the content delivery delay with ${\varepsilon {\rm{ = 5000}}}$ achieves a good trade-off between exploration of action space and exploitation of the result. As a result, the algorithm with ${\varepsilon {\rm{ = 5000}}}$ achieves lower converged content delivery delay, which also achieves a better convergence speed than the algorithm with ${\varepsilon {\rm{ = 9000}}}$ and ${\varepsilon {\rm{ = 13000}}}$.
This is a good proof of \textbf{Remark~\ref{remark3}}.
Fig.~\ref{fig5re2} shows the content delivery delay comparison with iteration numbers.
The performance of the QL based algorithm is evaluated with
different discount factors ${\gamma {\rm{ = \{ 0}}{\rm{.9}}, {\rm{0}}{\rm{.92}}, {\rm{0}}{\rm{.94\} }}}$.
It is observed from Fig.~\ref{fig5re2} that the content delivery delay of the QL based algorithm decreases and converges gradually with the number of iterations.
This is because the probability that the UAV choosing the optimal actions increases with iterations. Compared with the benchmark algorithms, the converged delay of the proposed algorithm is much smaller than that of the random algorithm, and approaches the greedy based exhaustive algorithm.

\begin{figure} [t!]
	\centering
	\includegraphics[width= 3.5in]{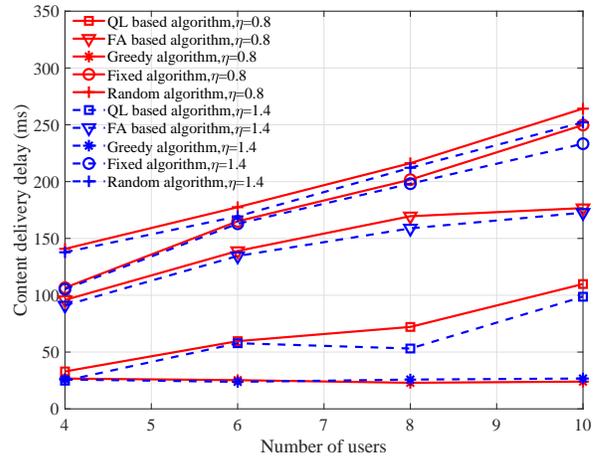}
	\caption{{Content delivery delay comparison with varying user numbers in small-scale scenario.}
	}
	\label{s2fig6re3}
\end{figure}

\begin{figure} [t!]
	\centering
	\includegraphics[width= 3.5in]{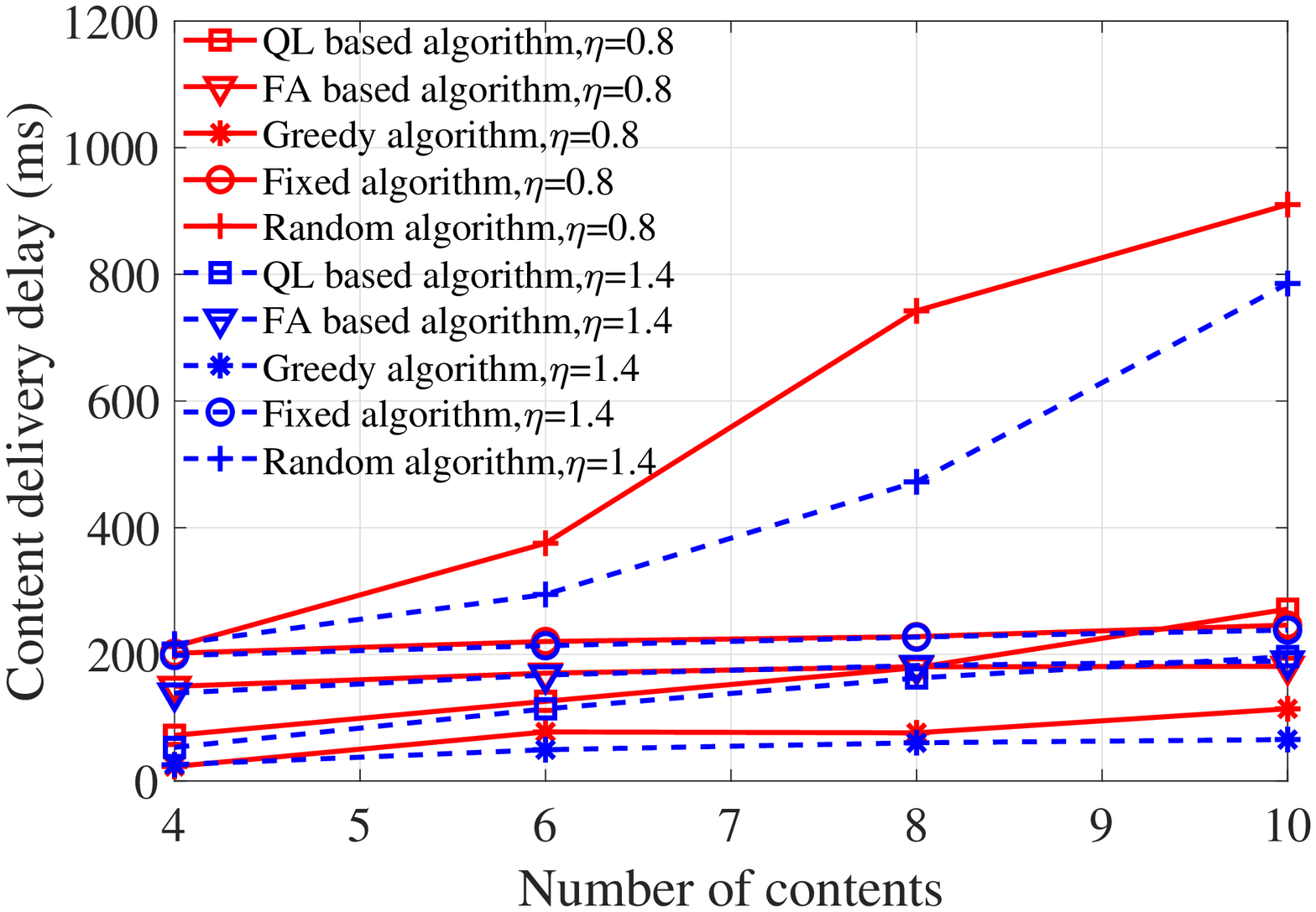}
	\caption{{Content delivery delay comparison with varying content numbers in small-scale scenario.}
	}
	\label{s2fig7re4}
\end{figure}

\begin{figure} [t!]
	\centering
	\includegraphics[width= 3.5in]{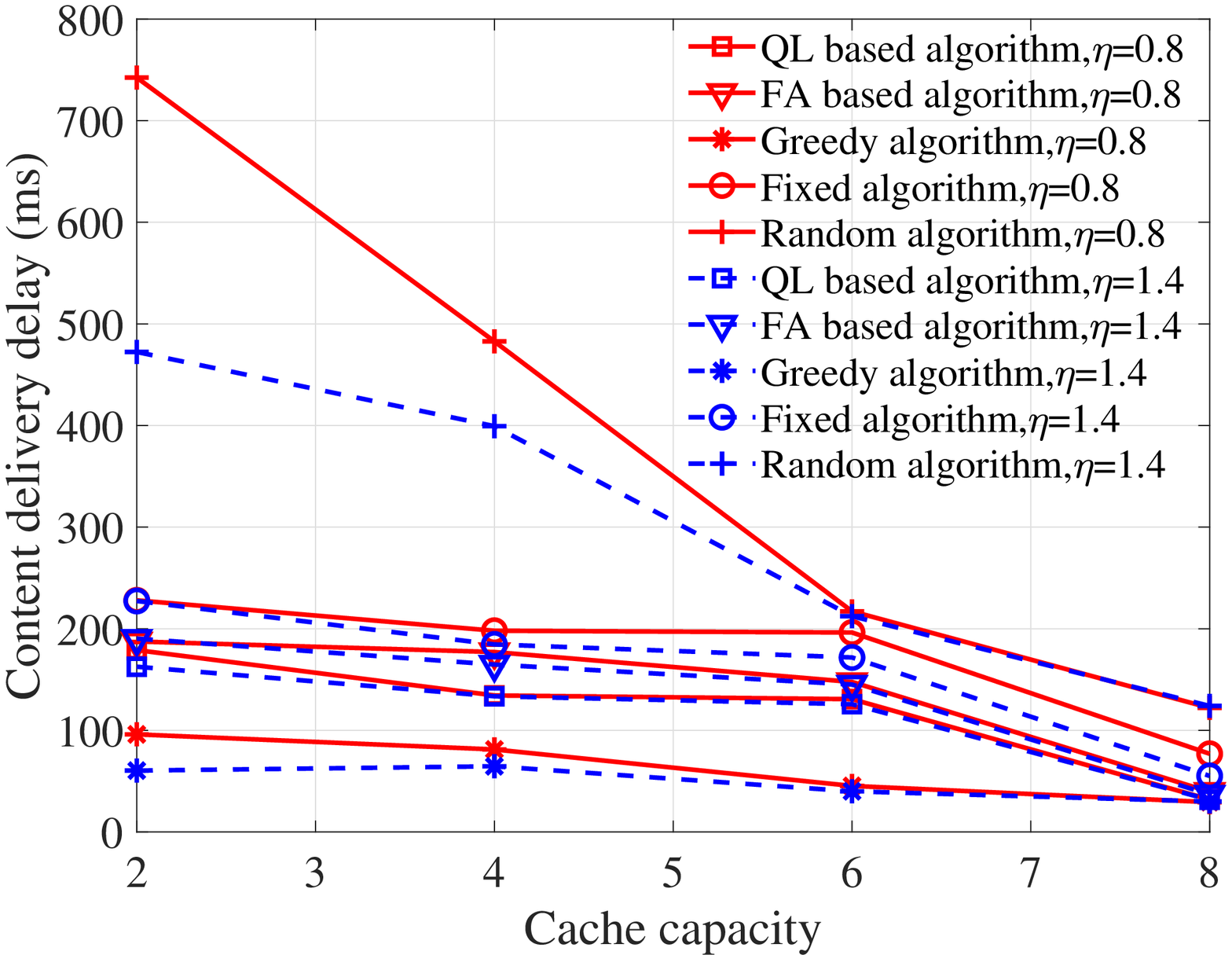}
	\caption{{Content delivery delay comparison with varying cache capacity in small-scale scenario.}
	}
	\label{s10fig10re7}
\end{figure}

\begin{figure} [t!]
	\centering
	\includegraphics[width= 3.5in]{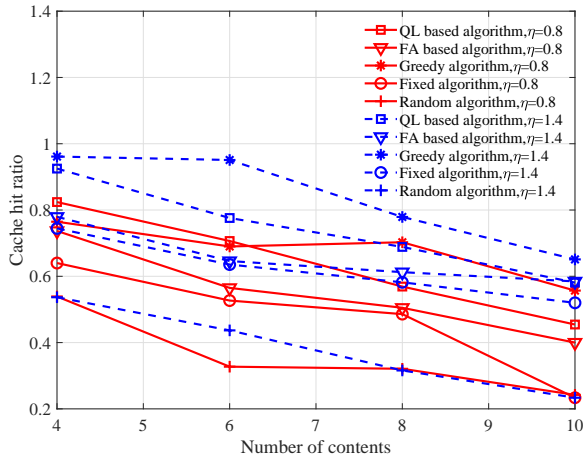}
	\caption{{Cache hit radio comparison with varying content numbers in small-scale scenario.}
	}
	\label{s10fig12re9}
\end{figure}

Then, we evaluate the proposed QL based algorithm and the proposed FA based algorithm in a small scale network with varying number of users, number of contents and cache capacity in Fig.~\ref{s2fig6re3}, Fig.~\ref{s2fig7re4}, and Fig.~\ref{s10fig10re7}, where ${\varepsilon {\rm{ = 5000}}}$ and ${\gamma  = 0.9}$.
The performance of the proposed algorithms and the benchmark algorithms are evaluated with different parameters of Zipf distribution ${\eta {\rm{ = }}\left\{ {{\rm{0}}{\rm{.8,1}}{\rm{.4}}} \right\}}$.
Fig.~\ref{s2fig6re3} shows the content delivery delay versus different numbers of users in the network with $M=4$, where
the content delivery delay increases monotonically with the number of users in the network.
Fig.~\ref{s2fig7re4} demonstrates the content delivery delay versus different numbers of contents in the network with ${N = 8}$, where the content delivery delay of the proposed algorithms increases with the number of contents. This is because that increase of number of the contents leads to bigger state space and legal action space, which reduces the probability that the optimal legal action is selected.
Fig.~\ref{s10fig10re7} demonstrates the content delivery delay versus different cache capacity in the network with ${N = 8}$ and ${M = 8}$, where the content delivery delay decreases with the cache capacity of the UAV.
As we can observe from both Fig.~\ref{s2fig6re3}, Fig.~\ref{s2fig7re4}, and Fig.~\ref{s10fig10re7}, the proposed algorithms achieve much smaller content delivery delay than the fixed algorithm and the random algorithm. The performance gap between the proposed algorithms and the greedy algorithm is relative small in the simulation.
However, the complexity of the proposed algorithms is much lower than that of the greedy algorithm especially when the size of the action space increases sharply with the number of users and contents increasing. Though there is a certain loss in the performance of the FA based algorithm compared to the QL based algorithm, the FA based algorithm is not limited by the action space and the state space. Thus the FA based algorithm could be deployed in large-scale networks.
Fig.~\ref{s2fig6re3}, Fig.~\ref{s2fig7re4}, and Fig.~\ref{s10fig10re7} also show that the parameter of the Zipf distribution slightly affects the algorithms when the numbers of contents and users are small.

Besides that, Fig.~\ref{s10fig12re9} demonstrates the cache hit ratio versus different numbers of contents in the network with ${N = 8}$, where the content delivery delay of the proposed algorithms decreases with the number of contents. As we can observe from Fig.~\ref{s10fig12re9}, the proposed algorithms achieve much higher cache hit radio than the fixed algorithm and the random algorithm. In some cases, the cache hit radio of QL based algorithm is better than greedy algorithm, this is because that the optimization objective is content delivery delay, which is not only determined by caching placement. Besides, the complexity of the proposed algorithms is much lower than that of the greedy algorithm.
\begin{figure} [t!]
	\centering
	\includegraphics[width= 3.5in]{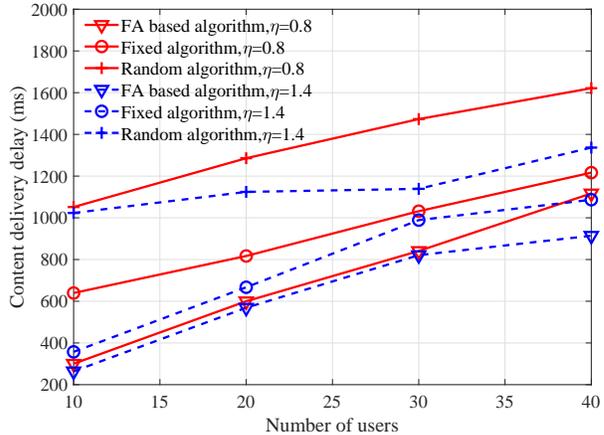}
	\caption{{Content delivery delay comparison with varying user numbers in large-scale scenario.}
	}
	\label{s3fig9re5}
\end{figure}

\begin{figure} [t!]
	\centering
	\includegraphics[width= 3.5in]{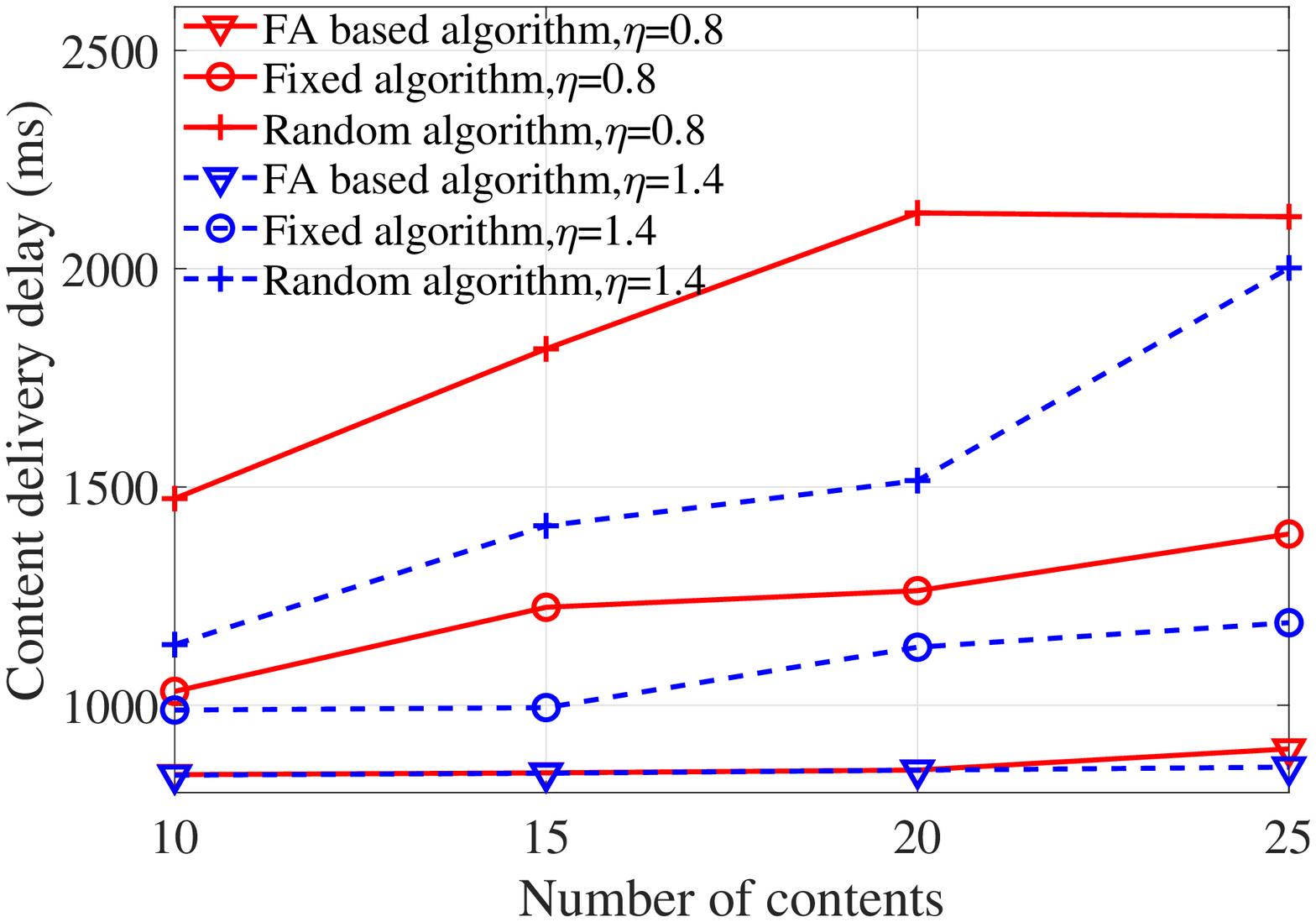}
	\caption{{Content delivery delay comparison with varying content numbers in large-scale scenario.}
	}
	\label{s3fig10re6}
\end{figure}

\begin{figure} [t!]
	\centering
	\includegraphics[width= 3.5in]{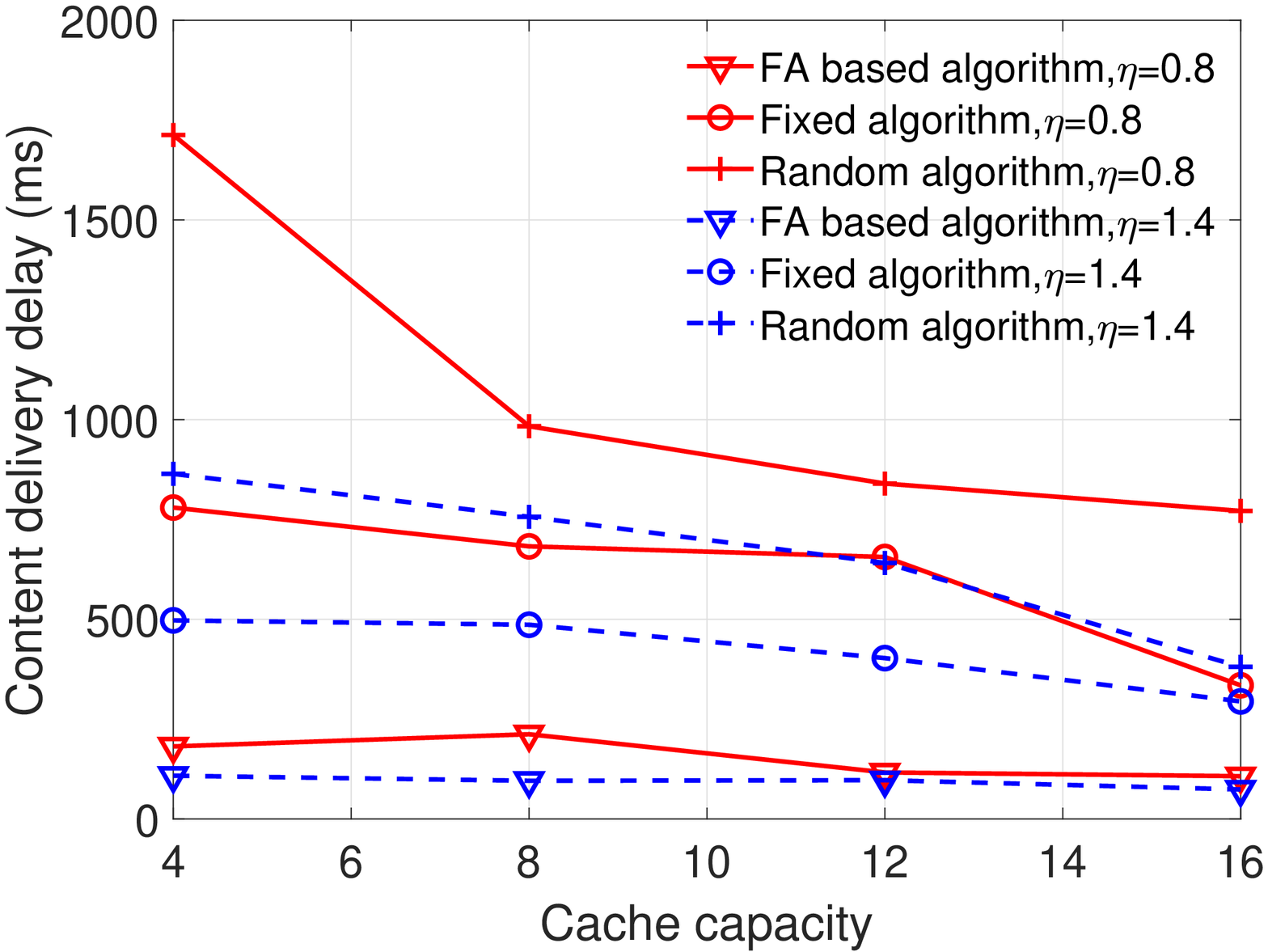}
	\caption{{Content delivery delay comparison with varying cache capacity in large-scale scenario.}
	}
	\label{s10fig11re8}
\end{figure}

\begin{figure} [t!]
	\centering
	\includegraphics[width= 3.5in]{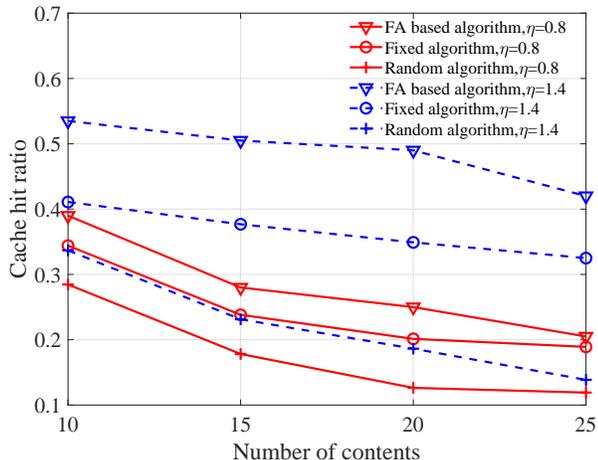}
	\caption{{Cache hit radio comparison with varying content numbers in large-scale scenario.}
	}
	\label{s10fig13re10}
\end{figure}



Since the size of action space increases greatly in large-scale networks, it is inefficient for the traditional Q-learning to search the action space. To deal with this, we use the FA based algorithm, which is compared with the benchmark algorithms in a large-scale network in Fig.~\ref{s3fig9re5}, Fig.~\ref{s3fig10re6}, and Fig.~\ref{s10fig11re8}.
The popularity of contents is generated according to Zipf distribution with different parameters ${\eta {\rm{ = }}\left\{ {{\rm{0}}{\rm{.8,1}}{\rm{.4}}} \right\}}$.
As the complexity of the greedy algorithm increases greatly in large-scale networks as well, we use the fixed algorithm and the random algorithm as the benchmark algorithms in the following simulation.
Fig.~\ref{s3fig9re5} reveals the performance of the FA based algorithm versus different numbers of users in the network with $M=10$, where the content delivery delay increases monotonically with the number of users in the network. The increase of user number has a certain impact on the performance of the algorithm, which verifies our obtained insights in \textbf{Remark~\ref{remark2}}. Fig.~\ref{s3fig10re6} reveals the performance of the FA based algorithm versus different numbers of contents with $N=30$, where the content delivery delay of the FA based algorithm increases with the number of contents. The increase in user number will enhance the performance advantages of the FA based algorithm relative to the fixed algorithm and the random algorithm, which verifies the insights from \textbf{Remark~\ref{remark1}}.
Fig.~\ref{s10fig11re8} reveals the performance of the FA based algorithm versus different cache capacity in the network with $N=30$ and $M=20$, where the content delivery delay decreases monotonically with the cache capacity in the network.
It is observed from Fig.~\ref{s3fig9re5}, Fig.~\ref{s3fig10re6}, and Fig.~\ref{s10fig11re8} that the FA based algorithm achieves much smaller content delivery delay than the fixed algorithm and the random algorithm. The reason is that the FA based algorithm can make more reliable decisions with efficient search mechanism.
Then, Fig.~\ref{s10fig13re10} demonstrates the cache hit ratio versus different numbers of contents in the network with ${N = 30}$, where the cache hit ratio of the proposed algorithm decreases with the number of contents. As we can observe from Fig.~\ref{s10fig13re10}, the proposed algorithm achieves much higher cache hit ratio than the fixed algorithm and the random algorithm.

Besides, compared with the content delivery delay and cache hit ratio of small-scale networks, parameters of Zipf distribution play a considerable role in the performance of the FA based algorithm in large-scale networks. This is because the concentrated users' interest distribution reduces the stress on DNN, which no longer needs to approximate users' unusual requests.

\section{Conclusion}
This article has investigated the cache-enabling UAV NOMA networks. The cache-enabling mobile UAV serves the user groups by NOMA and caches limited popular contents for wireless backhaul link traffic offloading. To model the uncertainty of dynamic environment, we have formulated the long-term caching placement and resource allocation optimization problem as a MDP. We have  defined the long-term sum delay of users as the content delivery cost, where the UAV acts as an agent.
The actions taken by UAV correspond to
caching placement, user scheduling and the power allocation of NOMA users.
We have used the QL-based algorithm and the FA-based algorithm to solve the dynamic optimization problem. Finally, numerical results show that the proposed algorithms yield significant performance gains compared to the fixed algorithm and the random algorithm, and have acceptable calculation complexity. Moreover, the results also show that the FA-based algorithm is not limited by the scale of networks.

\begin{appendices}
\section{Proof for NP-Hard of (\ref{optimal1})\label{appenA}}
Let ${{b_n}\left( t \right) = {r_{nm}}\left( t \right)}$ and ${{h_{NUg}}\left( t \right) = h_{NU}^1}$ to focus on the optimization on caching placement, then the proposed problem (\ref{optimal1}) could be conversed as
\begin{subequations}\label{optimal2}
	\begin{align}
	&\min_{b,i,h}  {\kern 1pt}{\kern 1pt}{\kern 1pt}{\sum\limits_{t = 1}^T {u\left( t \right)} }  \quad\\
	&{\kern 1pt}{\kern 1pt}{\kern 1pt}{\kern 1pt}{\kern 1pt}{\rm{s}}{\rm{.t}}{\rm{.}}{\kern 1pt} {\kern 1pt} {\kern 1pt} {\kern 1pt}{i_m}\left( t \right) \in \{ 0,1\} ,\forall m,t,\\
	\label{32d}	
	&{\kern 1pt}{\kern 1pt}{\kern 1pt} {\kern 1pt}{\kern 1pt}{\kern 1pt}{\kern 1pt}{\kern 1pt}{\kern 1pt}{\kern 1pt}{\kern 1pt}{\kern 1pt}{\kern 1pt}{\kern 1pt}{\kern 1pt}{\kern 1pt}{\kern 1pt}{\kern 1pt}{\kern 1pt}\sum\limits_{m = 1}^M {{i_m}\left( t \right)}  \le Z,\forall t,
	\end{align}
\end{subequations}
It is obvious that calculation complexity of problem (\ref{optimal2}) is not bigger than that of the formulated problem (\ref{optimal1}).
Since the problem (\ref{optimal2}) could be reduced to a 0-1 package problem with in polynomial-time~\cite{Tco7056418}, problem (\ref{optimal2}) is NP-hard. Hence, problem (\ref{optimal1}) is NP-hard.
\end{appendices}

\bibliographystyle{IEEEtran}
\bibliography{mybib}
%
\begin{IEEEbiography}[{\includegraphics[width=1in,height=1.25in,clip,keepaspectratio]{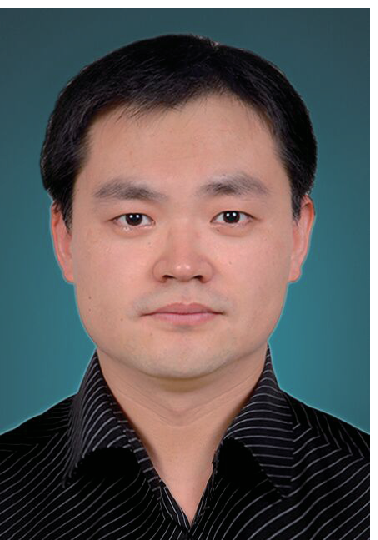}}]{Tiankui Zhang} (M'10-SM'15) received the Ph.D. degree in Information and Communication Engineering and B.S. degree in Communication Engineering from Beijing University of Posts and Telecommunications (BUPT), China, in 2008 and 2003, respectively. Currently, he is a Professor in School of Information and Communication Engineering at BUPT. His research interests include wireless communication networks, mobile edge computing and caching, signal processing for wireless communications, content centric wireless networks. He had published more than 100 papers including journal papers on IEEE Journal on Selected Areas in Communications, IEEE Transaction on Communications, etc., and conference papers, such as IEEE GLOBECOM and IEEE ICC.
\end{IEEEbiography}
\begin{IEEEbiography}[{\includegraphics[width=1in,height=1.25in,clip,keepaspectratio]{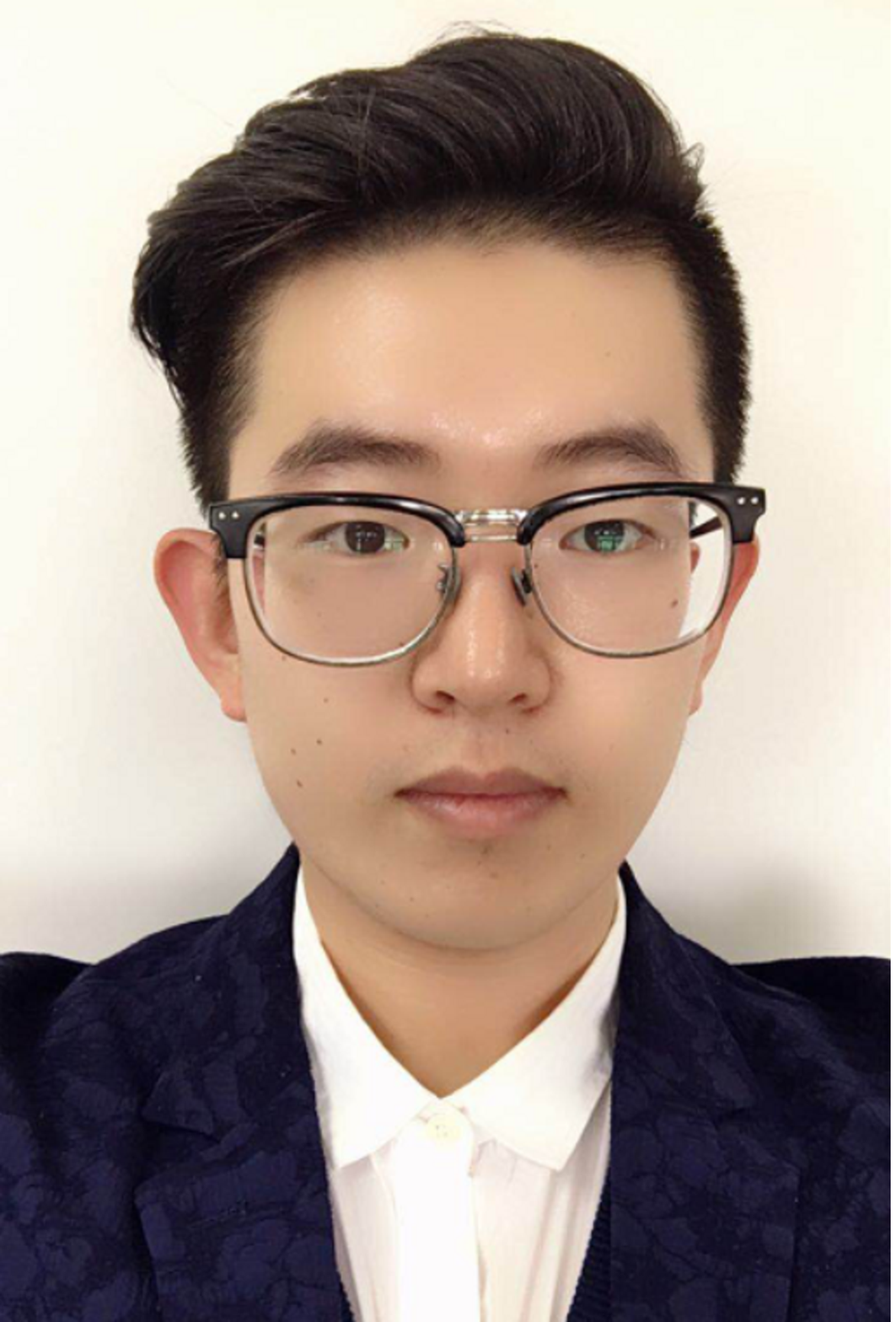}}]
	{Ziduan Wang} received the B.S. degree in Communication Engineering from  University of Electronic Science and Technology of China (UESTC) in 2018. He is currently working toward the M.S. degree in Information and Communication Engineering from Beijing University of Posts and Telecommunications (BUPT), China. His current research focuses on caching placement and resource allocation in cache-enabling UAV NOMA networks.
\end{IEEEbiography}


\begin{IEEEbiography}[{\includegraphics[width=1in,height=1.25in,clip,keepaspectratio]{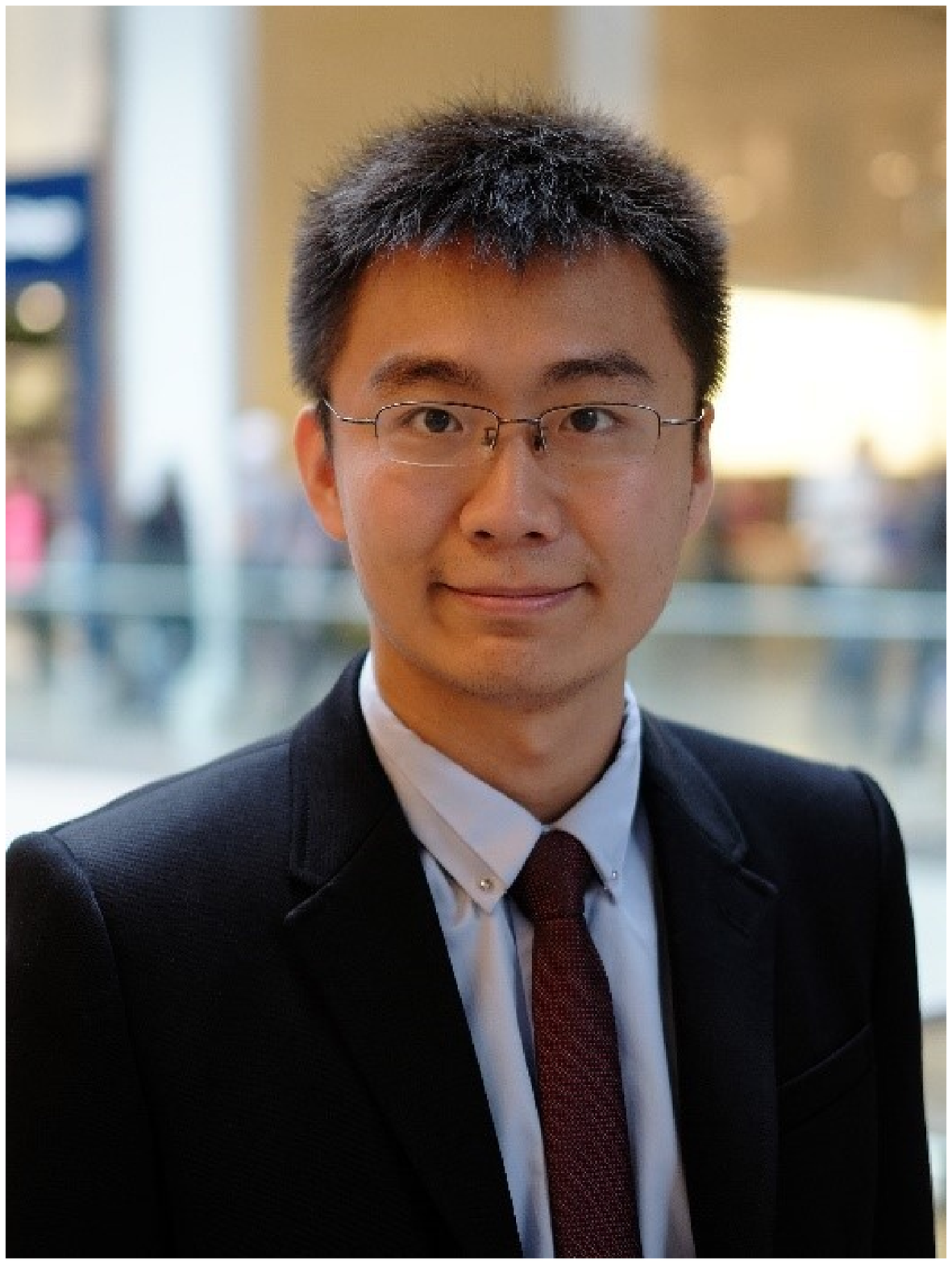}}] {Yuanwei Liu} (S'13-M'16-SM'19) received the B.S.  and M.S. degrees from the Beijing University of  Posts and Telecommunications in 2011 and 2014,  respectively, and the Ph.D. degree in electrical engineering from the Queen Mary University of London,  U.K., in 2016. He was with the Department of Informatics, King's  College London, from 2016 to 2017, where he  was a Post-Doctoral Research Fellow. He has been  a Lecturer (Assistant Professor) with the School  of Electronic Engineering and Computer Science, Queen Mary University of London, since 2017.
	
His research interests include  5G and beyond wireless networks, the Internet of Things, machine learning, and stochastic geometry. He has served as a TPC Member for many IEEE  conferences, such as GLOBECOM and ICC. He received the Exemplary Reviewer Certificate of IEEE WIRELESS COMMUNICATIONS LETTERS in  2015, IEEE TRANSACTIONS ON COMMUNICATIONS in 2016 and 2017, and IEEE TRANSACTIONS ON WIRELESS COMMUNICATIONS in 2017 and  2018. He has served as the Publicity Co-Chair for VTC 2019-Fall. He is currently an Editor on the Editorial Board of the IEEE TRANSACTIONS  ON COMMUNICATIONS, IEEE COMMUNICATIONS LETTERS, and IEEE ACCESS. He also serves as a Guest Editor for IEEE JSTSP special issue on Signal Processing Advances for Non-Orthogonal Multiple Access in Next Generation Wireless Networks.
\end{IEEEbiography}


\begin{IEEEbiography}[{\includegraphics[width=1in,height=1.25in,clip,keepaspectratio]{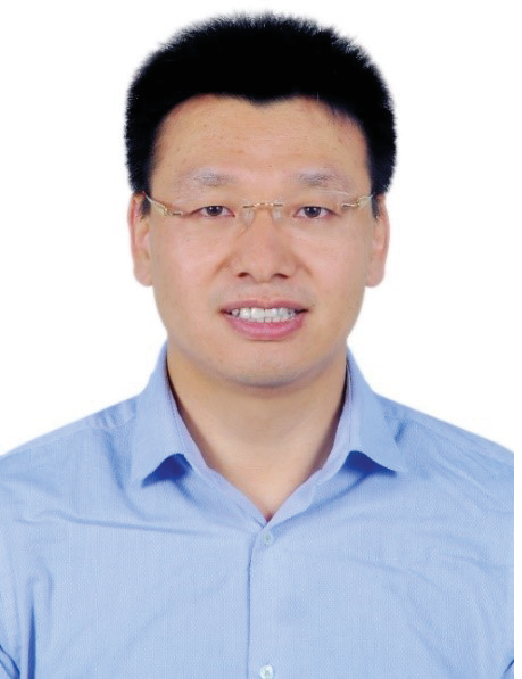}}]{Wenjun Xu} is a professor and Ph.D. supervisor in School of Information and Communication Engineering at Beijing University of Posts and Telecommunications (BUPT), Beijing, China. He received his B.S. and Ph.D. degrees from BUPT, in 2003 and 2008, respectively. He currently serves as a center director of the Key Laboratory of Universal Wireless Communications, Ministry of Education, P. R. China. He is a senior member of IEEE, and is now an Editor for China Communications. His research interests include AI-driven networks, UAV communications and networks, green communications and networking, and cognitive radio networks.
\end{IEEEbiography}


\begin{IEEEbiography}[{\includegraphics[width=1.0in,height=1.25in,clip,keepaspectratio]{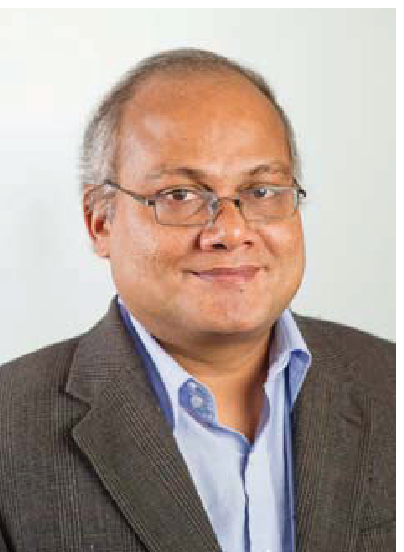}}]
{Arumugam Nallanathan} (S'97-M'00-SM'05-F'17) is Professor of Wireless Communications and Head of the Communication Systems Research (CSR) group in the School of Electronic Engineering and Computer Science at Queen Mary University of London since September 2017. He was with the Department of Informatics at Kings College London from December 2007 to August 2017, where he was Professor of Wireless Communications from April 2013 to August 2017 and a Visiting Professor from September 2017. He was an Assistant Professor in the Department of Electrical and Computer Engineering, National University of Singapore from August 2000 to December 2007. His research interests include Artificial Intelligence for Wireless Systems, Beyond 5G Wireless Networks, Internet of Things (IoT) and Molecular Communications. He published nearly 500 technical papers in scientific journals and international conferences. He is a co-recipient of the Best Paper Awards presented at the IEEE International Conference on Communications 2016 (ICC'2016) , IEEE Global Communications Conference 2017 (GLOBECOM'2017) and IEEE Vehicular Technology Conference 2018 (VTC'2018). He is an IEEE Distinguished Lecturer. He has been selected as a Web of Science Highly Cited Researcher in 2016 and an AI 2000 Internet of Things Most Influential Scholar in 2020.
	
He is an Editor-at-Large for IEEE Transactions on Communications and Senior Editor for IEEE Wireless Communications Letters. He was an Editor for IEEE Transactions on Wireless Communications (2006-2011), IEEE Transactions on Vehicular Technology (2006-2017) and IEEE Signal Processing Letters. He served as the Chair for the Signal Processing and Communication Electronics Technical Committee of IEEE Communications Society and Technical Program Chair and member of Technical Program Committees in numerous IEEE conferences. He received the IEEE Communications Society SPCE outstanding service award 2012 and IEEE Communications Society RCC outstanding service award 2014.
\end{IEEEbiography}

\end{document}